\documentclass{article}
\usepackage[utf8]{inputenc}

\usepackage{mathptmx}
\usepackage{adjustbox}
\usepackage{makecell}
\usepackage{upgreek} 
\usepackage{inconsolata}
\usepackage{amsmath}
\usepackage{breqn}
\usepackage{graphicx}
\usepackage{xcolor}
\usepackage{ragged2e}
\usepackage{subfig}
\usepackage{epstopdf}
\usepackage{xspace}
\usepackage{fancyvrb}
\usepackage{hyperref}
\usepackage{url}
\usepackage{url}
\usepackage{upgreek} 
\usepackage{amssymb}
\usepackage{listings}
\usepackage{soul}

\usepackage{geometry}
\newcommand{\rev}[1]{{#1}\xspace}
\newcommand{\ch}[1]{{#1}\xspace}

\newcolumntype{R}[2]{%
    >{\adjustbox{angle=#1,lap=\width-(#2)}\bgroup}%
    l%
    <{\egroup}%
}
\newcommand*\rot{\multicolumn{1}{R{90}{1em}}}

\usepackage{tikz}
\usetikzlibrary{arrows,shapes,positioning,shadows,trees}

\tikzset{
  basic/.style  = {draw, text width=7.5em, font=\sffamily, rectangle},
  root/.style   = {basic, rounded corners=4pt, thin, align=center,
                   fill=none, font=\sffamily},
  level 2/.style = {basic, rounded corners=6pt, thin,align=center, fill=white,
                   text width=9em, font=\small\sffamily},
  level 3/.style = {basic, thin, align=left, draw=none, fill=none, text width=7em, font=\footnotesize\sffamily}
}

\definecolor{mygreen}{rgb}{0,0.6,0}
\definecolor{mygray}{rgb}{0.5,0.5,0.5}

\lstdefinestyle{mystyle}{
    basicstyle=\footnotesize\ttfamily,
    frame=single,
    stepnumber=1,
    language=Prolog,
    keywords={chain, service, and, or, flow, maxLatency, node, link, placement, servicePlacement, flowPlacement, thingReqsOK, secReqsOK, hwReqsOK, path, update, updateOne, latencyOK, chainLatency, thingsReqsOK},
    keywordstyle=\color{blue}\ttfamily,       
    commentstyle=\it\color{mygreen}\ttfamily,
    numberstyle=\tiny\color{mygray}\ttfamily,
    rulecolor=\color{gray},
    breakatwhitespace=false,         
    breaklines=true,                 
    captionpos=b,                    
    keepspaces=true,
    numbers=left, 
     sensitive=true,
    numbersep=5pt,                  
    showspaces=false,                
    showstringspaces=false,
    showtabs=false,                  
    tabsize=2,
    breaklines=true
}

\newcommand{\problog}{{ProbLog}\xspace}
\newcommand{\prototype}{{{\small\textsf{EdgeUsher}}}\xspace}
\newcommand{\prototypefn}{{{\scriptsize\textsf{EdgeUsher}}}\xspace}

\title{Probabilistic QoS-aware Placement\\ of VNF chains at the Edge\thanks{This manuscript is a preprint version of the article published in \textit{Theory and Practice of Logic Programming}. The published article is available at: \url{http://doi.org/10.1017/S1471068421000016}}}
\author{S. Forti, F. Paganelli, A. Brogi}
\date{\small Department of Computer Science, University of Pisa, Italy}

\begin{document}

\maketitle

\begin{abstract}
\noindent
Deploying IoT-enabled Virtual Network Function (VNF) chains to Cloud-Edge infrastructures requires determining a placement for each VNF that satisfies all set deployment requirements as well as a software-defined routing of traffic flows between consecutive functions that meets all set communication requirements. 
In this article, we present a declarative solution, \prototype, to the problem of how to best place VNF chains to Cloud-Edge infrastructures. \prototype can determine all eligible placements for a set of VNF chains to a Cloud-Edge infrastructure so to satisfy all of their hardware, IoT, security, bandwidth, and latency requirements. It exploits probability distributions to model the dynamic variations in the available Cloud-Edge infrastructure, and to assess output eligible placements against those variations.

  \end{abstract}
  
{\footnotesize
\tableofcontents
}  
\section{Introduction}
\label{sec:introduction}

New Edge computing~\cite{abbas2018mobile} infrastructures aim at supporting Internet of Things (IoT) applications, especially when applications must meet stringent Quality of Service (QoS) requirements (e.g. latency, bandwidth, security) or handle large amounts of data. To achieve this goal, such new distributed infrastructures rely on computing capabilities which are closer to the edge of the Internet and to where data is produced and consumed (e.g. personal devices, access-points, smart network gateways, base stations, switches, and micro-datacentres). 
The fruitful interplay between Cloud and Edge resources aims at realising a Cloud-IoT continuum \cite{Puliafito2019}, also commonly known as Fog computing \cite{yousefpour2019all}. 

\rev{Meanwhile, the ongoing evolution of network technologies, namely Software-Defined Networking (SDN)} \cite{nguyen2017sdn} \rev{and Network Function Virtualisation (NFV)} \cite{mijumbi2015network}, \rev{is targeting a more effective usage of network resources for containing deployment and operational costs while coping with dynamic traffic demands, including requirements for delivering customised IoT-enabled services} \cite{baktir2017can}. 

\rev{On one hand, SDN decouples network \textit{control} \textit{plane} functions (e.g., routing decisions) from \textit{data} \textit{plane} functions (i.e., actual packet forwarding).
The SDN control plane can leverage a logically centralised management view of network resources to make informed control decisions, and the data plane API of network devices to fully program packet forwarding. Thus, SDN permits defining policies for more agile and cost-effective network operations, e.g. on-demand and QoS-aware traffic flow management, and optimised resource sharing/partitioning.}

On the other hand, NFV permits realising \rev{complex network services as chains of Virtual Network Functions} (VNFs), having each function \rev{(e.g. firewall, NAT, video optimiser)} implemented as a software application that can be installed and run on commodity hardware, virtual machines or containers. \rev{Each VNF can be seen as an independently deployable application service, which can be chained (viz. composed) with others so to apply their functionalities over data flows steered through them by means of SDN-enabled programmable forwarding devices.}

%
In this context, to fully release the potential of combining Edge and Cloud computing paradigms with new generation networks, novel models and methodologies should be devised in support of \textit{decision making} when deploying VNF chains that enable IoT applications \cite{alhussein2018joint}.
The problem of placing VNF chains consists of \textit{jointly} determining:

\vspace{-1.75mm}
\begin{itemize}
    \item[-] suitable \textit{VNF placements} over the available infrastructure \rev{(viz. mapping from each VNF to the node that will host it at run time)}, so to guarantee fulfilling all IoT (i.e., sensors and actuators), hardware and security requirements of each virtual service function, and
    \item[-] suitable \textit{routing of traffic flows} from one function to the other \rev{(viz. between the nodes that will host them at run time)}, so to satisfy end-to-end bandwidth and latency constraints. 
\end{itemize}

\vspace{-1mm}
\noindent Both decisions concern what is known as \textit{VNF placement} \cite{addis2015virtual}, or \textit{VNF chaining and embedding} \cite{alhussein2018joint}, and the corresponding problems are NP-hard \cite{sun2016forecast}. 
In order to solve these problems in Edge computing scenarios, new peculiarities of Edge infrastructures should be considered that were not considered in Cloud-only settings.

First and foremost, the edge of the Internet is characterised by the presence of resource-constrained (sometimes battery-powered) and heterogeneously capable devices, which communicate via wired and wireless communication technologies. This leads to a potentially high uncertainty and to \textit{variations in the available infrastructure} for what concerns both the availability of resources (e.g., due to node workload variations) and the QoS of the communication links (e.g., due to traffic variations). 

Second, as an extension to the Cloud, Edge computing will inherit from it 
many \textit{security threats}, while including its peculiar ones \cite{Farris2019}. On one hand, the number of security enforcement points will increase by allowing local processing of private data closer to the IoT sources. 
On the other hand, new infrastructures will have to face brand new threats for what concerns the physical vulnerability of devices. Indeed, VNF deployments to Edge infrastructures will include accessible (Edge or IoT) devices that may be easily hacked, stolen or broken by malicious users \cite{ni2017securing}.  
\rev{As highlighted by} \cite{Cziva2018}, most of the existing work on VNF placement in Cloud scenarios, as well as preliminary work targeting Edge computing, solved the problem only considering static infrastructure conditions.
For what concerns security aspects, to the best of our knowledge, \rev{only a few works} \cite{Fischer:2017,dwiardhika2019virtual,sendi2018} \rev{have been proposed that consider them when deciding on VNF placement and SDN-enabled routing in Cloud-Edge scenarios}\cite{Farris2019}.

In this article, we present a simple, yet general, probabilistic declarative methodology and a (heuristic) backtracking strategy to model and solve the \textit{VNF placement} problem in \rev{\textit{dynamic}} Cloud-Edge computing scenarios, \rev{while considering hardware, IoT, security, bandwidth and end-to-end latency requirements of the VNF chain to be deployed}. The methodology has been open-sourced by means of the probabilistic logic programming language \problog \cite{problog15} in the prototype \prototype\footnote{Available at: \url{https://github.com/di-unipi-socc/EdgeUsher}} and also allows to easily specifying and considering placement constraints (i.e. affinity and anti-affinity among functions). 
\rev{The main novel contribution of this work, exploiting a probabilistic declarative description of the VNF chain placement problem, is in that} \prototype\ \rev{permits determining VNF chain placements that are likely to ensure high QoS guarantees, security and service reliability over dynamic Cloud-Edge infrastructures that will be needed to achieve the ultimate NFV and SDN vision} \cite{laghrissi2018survey}.

\rev{Since it follows a \textit{declarative} implementation,} \prototype\ \rev{is more concise, easier to understand, modify and maintain with respect to procedural solutions, and it shows a high level of flexibility, being extensible and capable of accommodating the possibly evolving needs of Cloud-IoT scenarios. Besides,} \prototype\ \rev{is intrinsically \textit{explainable} as it derives proofs for input user queries by relying on state-of-the-art resolution engines, and it can be easily extended to justify \textit{why} a certain management decision was taken at runtime in the spirit of explainable AI .}

\rev{The rest of this article is organised as follows. After formulating the VNF chain placement problem and highlighting the main features of the proposed solutions} (Sect.~\ref{sec:problem}), \rev{the Prolog implementation of} \prototype\ \rev{is incrementally described by means of a series of examples} (Sect.~\ref{sec:methodology}). \rev{Then, the probabilistic heuristic version of} \prototype\ \rev{-- exploiting} \problog\ \rev{capabilities -- is described} (Sect.~\ref{sec:probabilistic_modelling}) \rev{and shown at work over a lifelike motivating example} (Sect.~\ref{sec:experiments}). \rev{A discussion of related work} (Sect.~\ref{sec:related}) \rev{and some lines for future work} (Sect.~\ref{sec:conclusions}) \rev{conclude the article\footnote{Appendix A lists the whole \prototypefn\ code and sketches a proof of termination and correctness of the proposed solving strategies.}.}


\vspace{-3mm}
\section{VNF Chain Placement: Problem Statement}
\label{sec:problem}

\rev{As aforementioned, the joint adoption of NFV and SDN technologies is considered by many as a promising approach to support next-gen IoT applications in Edge and Fog environments (Massonet et al. 2017; Farris et al. 2019). Indeed, those networking technologies are expected to enable the flexible matching of diverse IoT traffic requirements, ranging from low-latency and deployment costs minimisation (Wang et al. 2018; Leivadeas et al. 2019) to security mechanisms coping with threats in dynamic distributed environments (Farris et al. 2019; Puliafito et al. 2019).}

\rev{In this context, our work aims at contributing to solve the problem of placing VNF chains into VNF- and SDN-enabled Cloud-Edge infrastructures. Such a problem can be generally stated as follows:} 

\medskip
\fbox{
\begin{minipage}{0.95\textwidth}
{\em
\smallskip
 \rev{Let $C$ be a VNF chain with a set of deployment requirements $R_{D}$ on its composing VNFs and a set of communication requirements $R_{C}$ to suitably support communication flows between VNFs, and let $I$ be a distributed Cloud-Edge infrastructure made of computational nodes and communication links.}
 
 \smallskip
\rev{Eligible solutions to the VNF chain placement problem are joint mappings:}
 \begin{itemize}
     \item[(i)] \rev{from each VNF of $C$ to some infrastructure node in $I$, meeting all requirements in $R_{D}$, and}
     \item[(ii)] \rev{from each communication flow of $C$ to a valid routing path across the links of $I$,  connecting consecutive VNFs mapped as per (i), and meeting all requirements in $R_{C}$.}
 \end{itemize}
 
\rev{Eligible solutions can include nodes hosting multiple VNFs and links supporting multiple communication flows, if the involved nodes and link capacities can meet cumulative requirements of the VNFs and flows they have to support as per (i) and (ii), respectively.}}
\smallskip
\end{minipage}
}

\medskip\medskip\noindent
\rev{This work tackles the described problem following a probabilistic declarative methodology, based on the logic programming paradigm. As we will show throughout this article, the proposed} \prototype\ \rev{methodology and prototype input:}

\begin{itemize}
    \item[--] a description of one (or more) VNF chain(s) along with its (their) hardware, IoT, and security requirements (i.e. $R_{D}$), and minimum bandwidth, maximum end-to-end latency and security requirements between consecutive VNFs in the specified chain(s) (i.e. $R_{C}$), and 
    \item[--] a probabilistic description of the corresponding hardware, IoT, security, bandwidth and latency capabilities offered by the available Cloud-Edge infrastructure (i.e. $I$).
\end{itemize}

\vspace{-1mm}
\noindent \rev{Based on those,} \prototype\ \rev{outputs a ranking of all eligible solutions for the input instance of the VNF chain placement problem. Solutions include the mapping of each chain VNF to its deployment nodes and the routing of traffic flows between those deployment nodes.
The ranking of the eligible solutions considers how likely is a certain placement to satisfy all chain requirements as the infrastructure state (probabilistically) varies.}

\section{EdgeUsher Methodology}
\label{sec:methodology}

\rev{In this section, we incrementally describe our methodology to solve the VNF chain placement problem, by following the declarative implementation}\footnote{\rev{Prototype and examples code is also available at:} \url{https://github.com/di-unipi-socc/EdgeUsher} } \rev{of our Prolog prototype,} \prototype. \ch{Working increments of the prototype are assessed against small running examples, excerpted from a lifelike motivating example on a university video-surveillance distributed application, which we fully describe– and use to assess} \prototype \ch{– in Sect.} \ref{sec:experiments}. \ch{Such examples refer to a video surveillance application (consisting of various service functions) to be deployed on a campus network connecting various buildings and, more specifically, computing nodes with different hardware, IoT and security capabilities.}

\subsection{Matching a VNF to an Infrastructure Node}

\subsubsection{Hardware Requirements}

\rev{In the first place, we consider a single VNF to deployed to a single infrastructure node by matching only its hardware requirements. In this scenario, a VNF can be simply declared as in}

\begin{lstlisting}[]
service(FId, HWReqs).
\end{lstlisting}

\noindent \rev{where} \texttt{FId} \rev{is an identifier of the considered function and} \texttt{HWReqs} \rev{is an integer representing the amount of a generic hardware unit needed by the function to run properly.}

\rev{Specularly, an infrastructure node to match a VNF with can be easily declared as in}

\begin{lstlisting}[]
node(NodeId, HWCaps).
\end{lstlisting}

\noindent \rev{where} \texttt{NodeId} \rev{is an identifier of the considered node and} \texttt{HWCaps} \rev{is an integer representing the availability of hardware units at that node.}

\rev{In these very simple settings, the matching of a VNF with a node that can support its hardware requirements can be simply achieved as in}

\begin{lstlisting}[]
servicePlacement(FId,NId):- service(FId, HWReqs),
                            node(NId, HWCaps), 
                            hwReqsOK(HWReqs, HWCaps).
hwReqsOK(HWReqs, HWCaps) :- HWCaps >= HWReqs.
\end{lstlisting}

\noindent\textbf{Example. } \rev{Consider a service function \texttt{feature_extr} that analyses video frames streamed through it and extracts features for further analyses and requires $5$ hardware units to run properly.} 
\rev{It can be declared as}

\begin{lstlisting}[]
service(feature_extr, 5).
\end{lstlisting}

\noindent \rev{Now, consider an infrastructure made of the following two nodes (named after the building they are installed at):}

\begin{lstlisting}[]
node(studentCenter, 8).
node(briggsHall, 2).
\end{lstlisting}

\noindent \rev{Naturally, querying the predicate \texttt{servicePlacement(feature_extr, N)} will result in the possibility of deploying \texttt{feature_extr} to the \texttt{studentCenter} node, which features $8$ hardware units, but not to the \texttt{briggsHall} node, which only features $2$ hardware units. }
\begin{flushright}
$\square$
\end{flushright}

\subsubsection{IoT Requirements} 

\rev{We can extend the representation of a VNF by including its requirements in terms of IoT devices it needs to sense data from or act upon in response to cyber-physical triggers.}
\rev{By assuming that IoT devices can be uniquely identified by a symbolic name, we can now declare a VNF as in}

\begin{lstlisting}[]
service(FId, HWReqs, TReqs).
\end{lstlisting}

\noindent \rev{where \texttt{TReqs} is a list \texttt{[IoTId1, ..., IoTIdk]} of all the identifiers of the IoT devices -- sensors and actuators -- that need to be reachable by the deployment node of the function identified by \texttt{FId}.}

\rev{Analogously, we extend the representation of an infrastructure node as in}

\begin{lstlisting}[]
node(NodeId, HWCaps, TCaps).
\end{lstlisting}

\noindent \rev{where \texttt{TCaps} is the list of the identifiers of the IoT devices -- sensors and actuators -- that the node reaches out directly.}

\rev{The \texttt{servicePlacement/2} predicate defined before can be simply extended with a check on the IoT requirements as in}

\begin{lstlisting}[]
servicePlacement(FId,NId):- service(FId, HWReqs, TReqs),
                            node(NId, HWCaps, TCaps), 
                            hwReqsOK(HWReqs, HWCaps),
                            thingsReqsOK(TReqs, TCaps).
% hwReqsOK/2 definition
thingsReqsOK(TReqs, TCaps):- subset(TReqs, TCaps).
\end{lstlisting}

\noindent\textbf{Example. } \rev{Consider a service function \texttt{cctv\_driver} that requires $1$ hardware unit to run properly, and to directly reach a CCTV system identified by \texttt{video1}. }
\rev{It can be declared as}

\begin{lstlisting}[]
service(cctv_driver, 1, [video1]).
\end{lstlisting}

\noindent \rev{Now, consider an infrastructure made of the following two nodes:}

\begin{lstlisting}[]
node(parkingServices, 2, [video1]).
node(briggsHall, 2, []).
\end{lstlisting}

\noindent \rev{Naturally, querying the predicate \texttt{servicePlacement(cctv_driver, N)} will result in the possibility of deploying \texttt{cctv\_driver} to the \texttt{parkingServices} node, which reaches out the required CCTV system and features enough hardware resources. }
\begin{flushright}
$\square$
\end{flushright}

\vspace{-5mm}
\subsubsection{Security Requirements}

\begin{figure}[!ht]
\centering
\scalebox{0.75}{
\begin{tikzpicture}[
  level 1/.style={sibling distance=35mm},
  edge from parent/.style={-,draw},
  >=latex]

\node[root] {\textbf{EDGE \\SECURITY}}
  child {node[level 2] (c1) {\textbf{Virtualisation}}}
  child {node[level 2] (c2) {\textbf{Communications}}}
  child {node[level 2] (c3) {\textbf{Data}}}
  child {node[level 2] (c4) {\textbf{Physical}}}
  child {node[level 2] (c5) {\textbf{Other}}};

\begin{scope}[every node/.style={level 3}]
\node [below of = c1, xshift=15pt] (c11) {Access Logs};
\node [below of = c11] (c12) {Authentication};
\node [below of = c12] (c13) {Host IDS};
\node [below of = c13] (c14) {Process Isolation};
\node [below of = c14] (c15) {Permission Model};
\node [below of = c15] (c16) {Resource Usage Monitoring};
\node [below of = c16] (c17) {Restore Points};
\node [below of = c17] (c18) {User Data Isolation};

\node [below of = c2, xshift=15pt] (c21) {Certificates};
\node [below of = c21] (c22) {Firewall};
\node [below of = c22] (c23) {IoT Data Encryption};
\node [below of = c23] (c24) {Node Isolation Mechanims};
\node [below of = c24] (c25) {Network IDS};
\node [below of = c25] (c26) {Public Key Cryptography};
\node [below of = c26] (c27) {Wireless Security};

\node [below of = c3, xshift=15pt] (c31) {Backup};
\node [below of = c31] (c32) {Encrypted Storage};
\node [below of = c32] (c33) {Obfuscated Storage};

\node [below of = c4, xshift=15pt] (c41) {Access Control};
\node [below of = c41] (c42) {Anti-tampering Capabilities};

\node [below of = c5, xshift=15pt] (c51) {Audit};

\end{scope}

\foreach \value in {1,2,3,4,5,6,7,8}
  \draw[-] (c1.195) |- (c1\value.west);

\foreach \value in {1,...,7}
  \draw[-] (c2.195) |- (c2\value.west);

\foreach \value in {1,...,3}
  \draw[-] (c3.195) |- (c3\value.west);
  
\foreach \value in {1,...,2}
  \draw[-] (c4.195) |- (c4\value.west);
  
  \foreach \value in {1}
  \draw[-] (c5.195) |- (c5\value.west);
\end{tikzpicture}
}
\caption[]{Security capabilities for Edge computing.}
\label{taxonomy}
\end{figure}
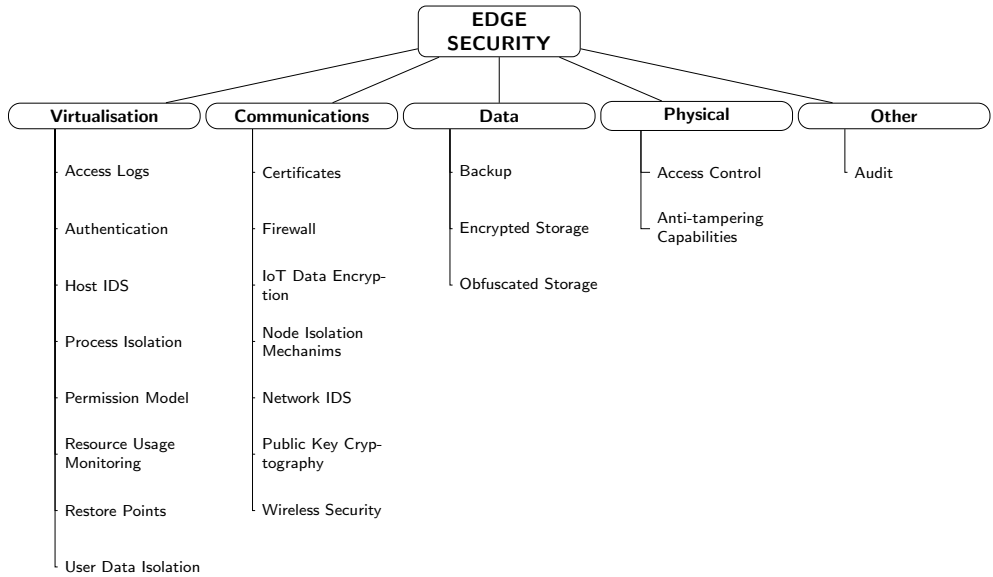

\rev{The logic-based formulation of the VNF placement problem that we are giving permits to naturally express security policies for a given VNF.
To this end, we assume that different infrastructure nodes can feature different security capabilities, expressed in terms of a common vocabulary of edge computing security capabilities, as per the taxonomy\footnote{\rev{The example taxonomy can be changed or extended so to include new security categories and third-level security capabilities as soon as normative security frameworks will get established in Cloud-Edge scenarios.}} of Fig.} \ref{taxonomy} \cite{fortisecure}.

\rev{Such a taxonomy can be used to express security policies for a given VNF as either a list or an AND/OR composition of security properties, over a common dictionary}.  

\medskip
\noindent\textbf{Example. } \rev{As an example, a security policy for a function collecting sensitive data at the edge of the Internet, could be expressed as the list}

\begin{lstlisting}[]
[anti_tampering, access_control]
\end{lstlisting}

\noindent
\rev{requiring both anti-tampering and access control capabilities to be available, as the deployment node can be easily accessed, stolen or broken by malicious users.}

\rev{As another example, a video processing functionality that stores and analyses possibly sensitive data for a long time might require enforcing the following AND/OR policy}

\begin{lstlisting}
and(access_control, or(obfuscated_storage, encrypted_storage))
\end{lstlisting}

\noindent\rev{which ensures the presence of an access control mechanism and of at least one data storage protection mechanism.}
\begin{flushright}
$\square$
\end{flushright}

\noindent
\rev{Building on top of this, we can now extend the representation of a VNF by including its security requirements as in}

\begin{lstlisting}[]
service(FId, HWReqs, TReqs, SecReqs).
\end{lstlisting}

\noindent \rev{where \texttt{SecReqs} is the security policy associated to the function identified by \texttt{FId}, represented as a list or an AND/OR composition of atomic security policies.}

\rev{Analogously, we extend the representation of an infrastructure node as in}

\begin{lstlisting}[]
node(NodeId, HWCaps, TCaps, SecCaps).
\end{lstlisting}

\noindent \rev{where \texttt{SecCaps} is the list of security capabilities featured by the node, expressed in terms of a common dictionary.}

\rev{Finally, the \texttt{servicePlacement/2} predicate defined before can be simply extended with a check on the security policies as in}

\begin{lstlisting}[]
servicePlacement(FId,NId):- service(FId, HWReqs, TReqs, SecReqs),
                            node(NId, HWCaps, TCaps, SecCaps), 
                            hwReqsOK(HWReqs, HWCaps),
                            thingsReqsOK(TReqs, TCaps),
                            secReqsOK(SecReqs, SecCaps).
                            
% hwReqsOK(HwReqs, HwCaps) definition
% thingsReqsOK(TReqs, TCaps) definition

secReqsOK([],_).                           
secReqsOK([SR|SRs], SecCaps)   :- subset([SR|SRs], SecCaps).
secReqsOK(and(P1,P2), SecCaps) :- secReqsOK(P1, SecCaps), 
                                  secReqsOK(P2, SecCaps).
secReqsOK(or(P1,P2), SecCaps)  :- secReqsOK(P1, SecCaps); 
                                  secReqsOK(P2, SecCaps).
secReqsOK(P, SecCaps) :- atom(P), member(P, SecCaps).
\end{lstlisting}

\noindent\textbf{Example. } \rev{Consider a service function \texttt{cctv\_driver} that requires $1$ hardware unit to run properly, to directly reach a CCTV system identified by \texttt{video1}, and the presence of either anti-tampering capabilities or an access control mechanism at the deployment node, or both. }
\rev{It can be declared as}

\begin{lstlisting}[]
service(cctv_driver, 1, [video1], or(anti_tampering, access_control)).
\end{lstlisting}

\noindent \rev{Now, consider an infrastructure made of the following two nodes that reach out the same CCTV system (\texttt{video1}) but feature different security capabilities:}

\begin{lstlisting}[]
node(parkingServices, 2, [video1], 
  [authentication, anti_tampering, wireless_security, obfuscated_storage]).
node(parkingServices2, 4, [video1], 
  [authentication, wireless_security, obfuscated_storage]).
\end{lstlisting}

\noindent \rev{Querying the predicate \texttt{servicePlacement(cctv_driver, N)} will result in the possibility of deploying \texttt{cctv\_driver} to the \texttt{parkingServices} node, as \texttt{parkingServices2} do not feature anti-tampering nor access control capabilities, despite satisfying all VNF requirements on hardware resources and IoT. }
\begin{flushright}
$\square$
\end{flushright}

\vspace{-6mm}
\subsection{Matching a VNF Chain to Infrastructure Nodes}
\label{sec:chaintoinfra}

\rev{At this stage, it is possible to easily specify chains of virtual network service functions and their hardware, IoT, network QoS and security requirements. Indeed, a VNF chain can be declared as}

\begin{lstlisting}[]
chain(ChainID, ServiceFunctionIDs).
\end{lstlisting}

\noindent
\rev{where \texttt{ChainID} uniquely identifies the chain and \texttt{ServiceFunctionIDs} lists the identifiers of all VNF composing the chain. }

\rev{In these settings, it is then possible to extend and exploit the \texttt{servicePlacement} predicate so to determine the placement of an entire chain, by placing one-by-one all services that compose it. This new behaviour is achieved by the \texttt{placement/2} predicate that follows }


\begin{lstlisting}[]
placement(Chain, Placement) :-
    chain(Chain, Services),
    servicePlacement(Services, Placement, []).

servicePlacement([], [], _).
servicePlacement([S|Ss], [on(S,N)|P], AllocatedHW) :-
    service(S, HWReqs, TReqs, SecReqs),
    node(N, HWCaps, TCaps, SecCaps),
    HW_Reqs =< HW_Caps,
    thingReqsOK(TReqs, TCaps),
    secReqsOK(SecReqs, SecCaps),
    hwReqsOK(HWReqs, HWCaps, N, AllocatedHW, NewAllocatedHW),
    servicePlacement(Ss, P, NewAllocatedHW).

% thingReqsOK(TReqs, TCaps) definition
% secReqsOk(SecReqs, SecCaps) definition

hwReqsOK(HWReqs, _, N, [], [(N, HWReqs)]).
hwReqsOK(HWReqs, HWCaps, N, [(N,A)|As], [(N,NewA)|As]) :-
    HWReqs + A =< HWCaps, NewA is A + HWReqs.
hwReqsOK(HWReqs, HWCaps, N, [(N1,A1)|As], [(N1,A1)|NewAs]) :-
    N \== N1, hwReqsOK(HWReqs, HWCaps, N, As, NewAs).
\end{lstlisting}

\rev{The new \texttt{servicePlacement/2} predicate inputs the list of {\tt Services} in the chain and returns an eligible {\tt Placement} of them to the available infrastructure. In doing so, not only it checks that hardware (line 9), IoT (line 10) and security requirements (line 11) of each VNF are satisfied but it also checks that cumulative hardware requirements of VNFs mapped to a same infrastructure node \texttt{N} do not exceed the overall capacity of the node. To this end, the Prolog program relies on an extended version of the \texttt{hwReqsOK} predicate (line 12, lines 18--22) to update the accumulator list \texttt{AllocatedHW} added as a third parameter in the \texttt{servicePlacement/3} predicate to keep track of the hardware resources as per the placement being built.}

\medskip
\noindent\textbf{Example. } \rev{As an example, consider a simple chain made of three VNFs that streams video footage from a CCTV system towards a feature extraction service function capable of identifying events of interests such as unauthorised vehicles access or fire and sending them to a lightweight analytics function for more accurate pattern recognition. Such a chain can be declared as in}

\begin{lstlisting}[]
chain(ucdavis_cctv, [cctv_driver, feature_extr, lw_analytics]).
service(cctv_driver, 1,[video1],
  or(anti_tampering,access_control)).
service(feature_extr, 3,[],
  and(access_control,or(obfuscated_storage,encrypted_storage))).
service(lw_analytics, 5,[],
  and(access_control,and(host_IDS,or(obfuscated_storage,encrypted_storage)))).
\end{lstlisting}

\medskip
\rev{Then, consider an infrastructure declaration with the following three nodes, featuring heterogeneous hardware, IoT and security capabilities:}

\begin{lstlisting}[]
node(parkingServices, 1, [video1], [authentication, anti_tampering,wireless_security,obfuscated_storage]).
node(westEntry, 1, [], [authentication, anti_tampering,wireless_security,obfuscated_storage]).
node(lifeSciences, 4, [video4], [access_logs, authentication, access_control, iot_data_encryption, firewall, host_IDS, pki, wireless_security, encrypted_storage]).
node(firePolice, 8, [video2, alarm1], [access_logs, access_control, authentication, backup,resource_monitoring, iot_data_encryption, firewall, host_IDS, pki, wireless_security, encrypted_storage]).
\end{lstlisting}

\rev{Querying the predicate \texttt{placement(cctv_driver, P)} will output the following placements
}

\vspace{1mm}
\begin{Verbatim}[fontfamily=courier, fontsize=\footnotesize, frame=single, framesep=1mm, framerule=0.1pt, rulecolor=\color{gray}]
placement(ucdavis_cctv,[on(cctv_driver,parkingServices), 
                        on(feature_extr,firePolice), 
                        on(lw_analytics,firePolice)])
                        
placement(ucdavis_cctv,[on(cctv_driver,parkingServices), 
                        on(feature_extr,lifeSciences), 
                        on(lw_analytics,firePolice)])
\end{Verbatim}

\noindent
\rev{where \texttt{cctv_driver} is always placed on the \texttt{parkingServices} so to reach out the required CCTV system, while \texttt{feature_extr} and \texttt{lw_analytics} can be placed either both on the \texttt{firePolice} node (which satisfies their cumulative hardware requirements of $8$ units), or on the \texttt{lifeSciences} and \texttt{firePolice} nodes, respectively.}

\begin{flushright}
$\square$
\end{flushright}

\vspace{-5mm}
\subsection{Routing Traffic Flows}
\label{sec:routing}

\rev{As a last step to complete the} \prototype\ \rev{prototype, we consider QoS requirements related to bandwidth allocation and end-to-end latency along a VNF chain. To this end, we extend the representation of a service with the information on its processing time as in}

\begin{lstlisting}[]
service(FId, TProc, HWReqs, IoTReqs, SecReqs).
\end{lstlisting}

\noindent
\rev{where \texttt{TProc} is the average time -- expressed in milliseconds -- elapsed between the instant an input is received by function \texttt{FId} and the instant the corresponding output is ready to be transmitted to the next function in the chain.}

\rev{We also extend the representation of a chain so to include the possibility of specifying bandwidth requirements as  directed traffic flows between couple of functions, as in}

\begin{lstlisting}[]
flow(FId1, FId2, BwReq).                 
\end{lstlisting}

\noindent \rev{where \texttt{FId1} and \texttt{FId2} are two VNF identifiers and \texttt{BWReq} is the bandwidth to be allocated via SDN directives along the path that connects their deployment nodes.}

\rev{Then, we permit specifying constraints on maximum tolerated latency for (directed) service paths crossing the functions} {\tt F1} $\rightarrow$ {\tt F2} $\rightarrow \cdots \rightarrow $ {\tt FN} \rev{as}

\begin{lstlisting}[]
maxLatency([F1, F2, ..., FN], LatReq).
\end{lstlisting}

\noindent
\rev{where \texttt{LatReq} is the end-to-end latency (in ms) not to be exceeded, summing up network and function processing delays.}

Finally, a (point-to-point or end-to-end) link\footnote{\prototypefn also permits specifying asymmetric links, for which upload and download QoS differ (e.g., like xDSL or 3/4G).} connecting {\tt NodeA} to {\tt NodeB} which is available in the considered infrastructure can be declared as

\begin{lstlisting}[]
link(NodeA, NodeB, Latency, Bandwidth).
\end{lstlisting}

\noindent where {\tt Latency} is the latency experienced over the link (in ms) and {\tt Bandwidth} is the transmission capacity it offers (in Mbps). 

The definition of the \texttt{placement} predicate illustrated before can be now extended so to check whether, for a given a \texttt{Placement} output by the service placement step, it is possible to determine eligible \texttt{Routes} for the chain traffic flows across function services. This is done by the \texttt{flowPlacement(Placement, ServiceRoutes)} predicate as in

\begin{lstlisting}[]
placement(Chain, Placement, Routes) :-
    chain(Chain, Services),
    servicePlacement(Services, Placement, []),
    flowPlacement(Placement, Routes).

% servicePlacement(S, P, AllHw) definition

flowPlacement(Placement, ServiceRoutes) :-
    findall(flow(S1, S2, Br), flow(S1, S2, Br), ServiceFlows),
    flowPlacement(ServiceFlows, Placement, [], ServiceRoutes, [], S2S_latencies),
    maxLatency(LChain, RequiredLatency),   
    latencyOK(LChain, RequiredLatency, S2S_latencies).

flowPlacement([], _, SRs, SRs, Lats, Lats).
flowPlacement([flow(S1, S2, _)|SFs], P, SRs, NewSRs, Lats, NewLats) :-
    subset([on(S1,N), on(S2,N)], P),
    flowPlacement(SFs, P, SRs, NewSRs, [(S1,S2,0)|Lats], NewLats). 
flowPlacement([flow(S1, S2, Br)|SFs], P, SRs, NewSRs, Lats, NewLats) :-
    subset([on(S1,N1), on(S2,N2)], P), N1 \== N2,
    path(N1, N2, 2, [], Path, 0, Lat),
    update(Path, Br, S1, S2, SRs, SR2s),
    flowPlacement(SFs, P, SR2s, NewSRs, [(S1,S2,Lat)|Lats], NewLats). 

path(N1, N2, Radius, Path, [(N1, N2, Bf)|Path], Lat, NewLat) :-
    Radius > 0, link(N1, N2, Lf, Bf), NewLat is Lat + Lf.
path(N1, N2, Radius, Path, NewPath, Lat, NewLat) :-
    Radius > 0, link(N1, N12, Lf, Bf), N12 \== N2, \+ member((N12,_,_,_), Path),
    NewRadius is Radius-1, Lat2 is Lat + Lf,
    path(N12, N2, NewRadius, [(N1, N12, Bf)|Path], NewPath, Lat2, NewLat).

update([],_,_,_,SRs,SRs).
update([(N1, N2, Bf)|Path], Br, S1, S2, SRs, NewSRs) :-
    updateOne((N1, N2, Bf), Br, S1, S2, SRs, SR2s),
    update(Path, Br, S1, S2, SR2s, NewSRs).

updateOne((N1, N2, Bf), Br, S1, S2, [], [(N1, N2, Br, [(S1,S2)])]) :-
    Br =< Bf.
updateOne((N1, N2, Bf), Br, S1, S2, [(N1, N2, Bass, S2Ss)|SR], [(N1, N2, NewBa, [(S1,S2)|S2Ss])|SR]) :- 
    Br =< Bf-Bass, NewBa is Br+Bass.
updateOne((N1, N2, Bf), Br, S1, S2, [(X, Y, Bass, S2Ss)|SR], [(X, Y, Bass, S2Ss)|NewSR]) :-
    (N1 \== X; N2 \== Y),
    updateOne((N1, N2, Bf), Br, S1, S2, SR, NewSR).
    
latencyOK(LChain, RequiredLatency, S2S_latencies) :-
    chainLatency(LChain, S2S_latencies, 0, ChainLatency),
    ChainLatency =< RequiredLatency.

chainLatency([S], _, Latency, NewLatency) :-
    service(S, S_Service_Time, _, _, _),
    NewLatency is Latency + S_Service_Time.
chainLatency([S1,S2|LChain], S2S_latencies, Latency, NewLatency) :-
    member((S1,S2,Lf), S2S_latencies),
    service(S1, S1_Service_Time, _, _, _),
    Latency2 is Latency+S1_Service_Time+Lf,
    chainLatency([S2|LChain], S2S_latencies, Latency2, NewLatency).
\end{lstlisting}

The program above first checks bandwidth requirements (line 10) and, afterwards, latency requirements (line 11--12).
First, a routing satisfying bandwidth constraints is determined by the predicate {\tt flowPlacement} (line 10) which holds if:

\vspace{-1.75mm}
\begin{itemize}
    \item[--] the services {\tt S1} and {\tt S2} in between which a flow is established have been placed onto the same node {\tt N} (lines 15--17), or
    \item[--] the services {\tt S1} and {\tt S2} in between which the flow is established have been placed onto different nodes {\tt N1} and {\tt N2} and there exists a path in between those nodes that supports the bandwidth requirement of the flow (lines 18--22).
\end{itemize}

\noindent The {\tt path(N1, N2, Radius, [], Path, 0, NewLat)} predicate determines an acyclic {\tt Path} of length at most {\tt Radius} (i.e., maximum hop number) in between {\tt N1} and {\tt N2}, which features latency {\tt Lat} (line 20). A path is either a direct infrastructure link between {\tt N1} and {\tt N2} (lines 24--25), or a route of links that connect them (lines 26--29). \rev{It is worth noting that even when setting the value of {\tt Radius} to low values $K$ (i.e., $2-3$) the found routing will actually be able to spread a chain of length $L$ over a path of length $K\times L$, thus extending the chain potential reach.    Naturally, it is possible to relax the constraint on the {\tt Radius} -- incurring in longer execution times -- by setting {\tt Radius} to values larger than the default one (viz. $2$) at line $20$.}

After a path is found, {\tt update} checks if the bandwidth requirements of the considered flow can be supported by such path (lines 31--34). Similarly to hardware allocation, a list of elements of the form {\tt (N1, N2, Bf)} is maintained to keep track of the bandwidth {\tt Bf} allocated on each link along a certain path and to check whether more flows mapped onto the same link do not exceed its capacity. Particularly, {\tt updateOne} scans the list of links along a path and checks such requirements by accumulating the bandwidth consumed by all flows routed onto the same link (lines 36--42).

Finally, {\tt latencyOK} holds if the chain latency -- which is computed by summing the functions processing times of the traversed functions with the latency of the chosen path (lines 48--55) -- is less than or equal to the one required by the specified {\tt maxLatency} requirement.

\medskip
\noindent\textbf{Example. } \rev{As an example, consider the chain of the previous example, extended with the following requirements on traffic flows and end-to-end latency}

\begin{lstlisting}[]
flow(cctv_driver, feature_extr, 15).                        
flow(feature_extr, lw_analytics, 8).

maxLatency([cctv_driver, feature_extr, lw_analytics], 50).
\end{lstlisting}

\rev{Then, consider the previous infrastructure declaration completed with the following links:}

\begin{lstlisting}[]
link(parkingServices, westEntry, 15, 70).
link(westEntry, parkingServices, 15, 70).

link(parkingServices, lifeSciences, 15, 70).
link(lifeSciences, parkingServices, 15, 70).

link(westEntry, firePolice, 15, 70).
link(firePolice, westEntry, 15, 70).
\end{lstlisting}

\rev{Querying the predicate \texttt{placement(cctv_driver, P, R)} will output the following placement and associated routing directives
}

\vspace{1mm}
\begin{Verbatim}[fontfamily=courier, fontsize=\footnotesize, frame=single, framesep=1mm, framerule=0.1pt, rulecolor=\color{gray}]
placement(ucdavis_cctv,
        [on(cctv_driver,parkingServices), 
         on(feature_extr,firePolice), 
         on(lw_analytics,firePolice)],
        [(westEntry, firePolice, 15, [(cctv_driver, feature_extr)]), 
         (parkingServices, westEntry, 15, [(cctv_driver, feature_extr)])])
\end{Verbatim}

\noindent
\rev{where \texttt{cctv_driver} is placed on the \texttt{parkingServices} and \texttt{feature_extr} and \texttt{lw_analytics} can only be placed on the \texttt{firePolice} node. Indeed, the previously determined placement of \texttt{feature_extr} to the \texttt{lifeSciences} node is not eligible anymore as the path that connects the \texttt{lifeSciences} node to the \texttt{firePolice} node cannot support the end-to-end latency of $50$ ms. It is worth noting that the traffic flow of $15$ Mbps between \texttt{cctv_driver} and \texttt{feature_extr} follows a path passing through \texttt{westEntry}, which connects \texttt{parkingServices} to \texttt{firePolice}. On the other hand, as \texttt{feature_extr} and \texttt{lw_analytics} are mapped onto the same node, no routing is output for the $8$ Mbps traffic flow in between them.}

\begin{flushright}
$\square$
\end{flushright}

\subsection{(Anti-)Affinity Constraints and Partial Solutions}
\label{sec:queries}

It is worth noting that \prototype allows users to easily specify placement constraints in the form of function affinity or anti-affinity requirements among functions. \rev{Affinity consists in placing two or more functions in the same physical node, thus reducing latency and communications costs between VNFs, while anti-affinity prevents two or more VNFs from sharing the same resources} \cite{oechsner2015flexible}. \rev{The possibility to add affinity and anti-affinity constraints is useful since it allows specifying deployment location requirements needed for performance, economic, resilience, legislative and privacy issues} \cite{BoutenCMFLS16}.

In the case of affinity constraints, the user can force the mapping of two (or more) functions to the same node, as for instance in the query

\begin{lstlisting}
placement(Chain, [on(F1,N1), on(F2,N2), on(F3,N2)], Routes).
\end{lstlisting}

\noindent stating that 
{\tt F2} and {\tt F3} must be mapped on a same node {\tt N2}.
Analogously, anti-affinity constraints can be specified by queries of the form:

\begin{lstlisting}
placement(Chain, [on(F1,N1), on(F2,N2), on(F3,N3)], Routes), N2 \== N3).
\end{lstlisting}

\noindent imposing that {\tt F2} and {\tt F3} must be mapped on two different nodes {\tt N2} and {\tt N3}.


Finally, users can specify partial deployments and/or routes, and use \prototype to complete them. This is useful to quickly determine on-demand re-configurations of part of a chain in case this is affected by infrastructure failures or malfunctioning (e.g., crash of a node hosting a function service) \rev{without recomputing (and eventually migrating) the whole chain}.

\subsection{Complexity Analysis}
\label{sec:complexity_analysis}

\prototype relies on logic programming backtracking mechanism to determine \rev{the} eligible VNF placement(s) and traffic routing(s) for a VNF chain to be deployed to an Edge infrastructure. 

The \rev{worst-case} time complexity of \texttt{servicePlacement} is clearly $O(n^{s})$, where $n$ is the number of nodes in the available infrastructure and $s$ is the number of services in the input service chain. 
The worst-case time complexity of \texttt{flowPlacement} is $O(b^{\tt Radius})$, where $b$ is the average out-degree of nodes in the infrastructure, and in any case it is bounded by $O(n^{\tt Radius})$ (\rev{when dealing with fully connected network topologies}).
\rev{These hold both in the case we aim at determining a \textit{single} eligible placement or routing (and only one exists, and it is the last one found via backtracking), and in the case in which we aim at determining \textit{all} eligible placements or routings (which naturally requires to fully explore the placement and routing search spaces, independently of the number of existing eligible solutions).}

\rev{
Now, the worst-case case for determining an eligible placement and routing happens when a single eligible solution exists and it corresponds to combining the last found placement with the last found routing. In such a case, the combination of \texttt{servicePlacement} with \texttt{flowPlacement} incurs in a time complexity of $O(n^s \times n^{\tt Radius}) = O(n^{s+{\tt Radius}})$. As per the above considerations, the worst-case time complexity of the approach described until now is exponential, and so is the exhaustive exploration of the search space. }

%
%

\section{Probabilistic Modelling}
\label{sec:probabilistic_modelling}

\rev{In this section, we first recapitulate on the probabilistic logic programming language} \problog\ \rev{(Sect.} \ref{sec:problog}). 
\rev{Then, we illustrate how the probabilistic capabilities of} \problog\ \rev{permit enhancing the Prolog} \prototype\ \rev{prototype} \rev{so to consider dynamic infrastructure conditions when solving the VNF embedding problem (Sect.} \ref{sec:piedgeusher}). 
\rev{Besides, we will show how} \problog\ \rev{meta-reasoning capabilities can be exploited to reduce the exp-time complexity discussed before (Sect.} \ref{sec:heuristics}).

\subsection{Background: The ProbLog Language}
\label{sec:problog}

Probabilistic logic programming extends logic programming by enabling the representation of uncertain information. More specifically, logic programming allows representing relations among entities, while probability theory can model uncertainty over attributes and relations \cite{riguzzi2018foundations}.
To implement both the model and the matching strategy we used  the \problog\ language \cite{kimmig2011implementation,problog15}, 
a probabilistic extension of Prolog.

\medskip
\noindent 
Prolog programs are finite sets of rules of the form

\vspace{1mm}
\begin{Verbatim}[fontfamily=courier, fontsize=\footnotesize, frame=single, framesep=1mm, framerule=0.1pt, rulecolor=\color{gray}]
a :- b1, ... , bn.
\end{Verbatim}

\noindent 
stating that {\tt a} holds when {\tt b1} $\wedge\ \cdots\ \wedge$ {\tt bn} holds, 
where {\tt n}$\geq$0 and {\tt a}, {\tt b1}, ..., {\tt bn} are atomic literals.  
Rules with empty conditions ({\tt n}$=$0) are also called facts.

\medskip
\noindent
\problog\ programs are logic programs in which some of the facts are annotated with probabilities. 
A \problog\ fact, such as 

\vspace{1mm}
\begin{Verbatim}[fontfamily=courier, fontsize=\footnotesize, frame=single, framesep=1mm, framerule=0.1pt, rulecolor=\color{gray}]
p::a.
\end{Verbatim}

\noindent states that  {\small\tt a} holds with probability {\small\tt p}.
Non-annotated facts are assumed to always hold with probability $1$.

\medskip
\noindent
Problog also allows to use semicolons to express OR conditions in rules. For instance

\begin{Verbatim}[fontfamily=courier, fontsize=\footnotesize, frame=single, framesep=1mm, framerule=0.1pt, rulecolor=\color{gray}]
a :- b1; ... ; bn.
\end{Verbatim}

\noindent states that {\tt a} holds when {\tt b1} $\vee\ \cdots\ \vee$ {\tt bn} holds. 

\medskip
\noindent
Finally, annotated disjunctions, like

\begin{Verbatim}[fontfamily=courier, fontsize=\footnotesize, frame=single, framesep=1mm, framerule=0.1pt, rulecolor=\color{gray}]
p1::a1; p2::a2; ...; pK::aK.
\end{Verbatim}

\noindent state 
 that at most one of the facts {\tt a1}, ..., {\tt aK} holds with the associated probability\footnote{If $T=\sum_{{\tt i}=1}^{\tt K} {\tt pi} < 1$, \problog assumes the presence of an implicit null choice which states with probability $1-T$ that none of the  {\tt K} options holds.}.

\medskip\noindent
Each \problog program defines a probability distribution over logic programs where a fact {\tt p::a.} is considered true with probability {\tt p} and false with probability $1 - ${\tt p}. 
The \problog\ engine \cite{problog15} determines the success probability of a query {\tt q} as the probability that {\tt q} has \textit{a} proof, given the distribution over logic programs. 

\rev{Intuitively, a} \problog\ \rev{program leverages input probability distributions to analyse all possible Prolog programs (i.e., worlds) that could be generated according to them. 
Assuming that $\Omega(q)$ is the set of possible worlds $W$ that entail a valid proof for a certain query $q$ 
(i.e., $\Omega(q) = \{W \mid W \models q\}$), the} \problog\ \rev{engine computes the probability $p(q)$ that $q$ holds as}
$$p(q) \ = \ \sum_{\substack{W \in \Omega(q)}} \ \prod_{f \in W} \ p(f)$$

\noindent \rev{where $f$ are facts within a possible world, and $p(f)$ is the probability they are labelled with.}

\subsection{Probabilistic {\textsf{EdgeUsher}}}
\label{sec:piedgeusher}

\rev{The usage of} \problog\ \rev{permits to naturally specify probabilistic profiles of both nodes and links by exploiting \textit{annotated} \textit{disjunctions}. Such language constructs permits to capture the intrinsic dynamicity and uncertainty of Edge infrastructures by relying on probability distributions based on historically monitored data. 

An infrastructure node can then be declared as in}

\vspace{-3mm}
\begin{lstlisting}
P1::node(NId, HwP1, IoTCapsP1, SecCapsP1);
P2::node(NId, HWP2, IoTCapsP2, SecCapsP2); 
...; 
Pk::node(NId, HWPk, IoTCapsPk, SecCapsPk).
\end{lstlisting}

\noindent \rev{where $\sum_{i = 0}^{\texttt{k}} \texttt{Pi} \leqslant 1$ and \texttt{Pi} is the probability that a particular node configuration (with respect to available hardware, IoT devices, and security capabilities) occurs.}

\rev{Analogously, links between nodes can be specified as in}

\vspace{-3mm}
\begin{lstlisting}
P1::link(N1, N2, LatP1, BwP1);
P2::link(N1, N2, LatP2, BwP2);
...;
Pk::link(N1, N2, LatPk, BwPk)
\end{lstlisting}

\noindent \rev{where $\sum_{i = 0}^{\texttt{k}} \texttt{Pi} \leqslant 1$ and \texttt{Pi} is the probability that a particular link QoS configuration (with respect to end-to-end latency and available bandwidth) occurs.
}

\rev{It is worth noting that three different types of facts (and any mixture of them) can constitute the input infrastructure description in the} \problog\ \rev{version of} \prototype. \rev{Indeed, a user can exploit:}
\begin{itemize}
    \item[(a)] \rev{\textit{non-probabilistic} facts when using real-time monitoring or averaged monitoring data (i.e., avoiding annotating facts), as shown in Sect. }\ref{sec:methodology},
    \item[(b)] \rev{a \textit{single-probability} facts that are just annotated with an indication of their reliability (e.g., \texttt{0.99::(cloudX, 30, [], [firewall]).}), or}
    \item[(c)] \rev{a \textit{fully} \textit{probabilistic} facts annotated with complete probability distributions describing the infrastructure dynamics based on aggregate historical monitoring data.}
\end{itemize}

\medskip\noindent
\rev{Naturally, running} \prototype\ in \problog\ \rev{ with probabilistic input enhances output precision via ranking eligible VNF placements and traffic routings according to how well they are expected to satisfy the chain requirements as the infrastructure state (probabilistically) varies. }

\medskip
\noindent\textbf{Example. } \rev{As an example, consider the chain of the previous example, to be matched to the following probabilistic infrastructure description}

\begin{lstlisting}[]
0.2::node(parkingServices, 2, [video1], [authentication, anti_tampering,wireless_security,obfuscated_storage]);
0.8::node(parkingServices, 1, [video1], [authentication, anti_tampering,wireless_security,obfuscated_storage]).

0.2::node(westEntry, 2, [], [authentication, anti_tampering,wireless_security,obfuscated_storage]);
0.8::node(westEntry, 1, [], [authentication, anti_tampering,wireless_security,obfuscated_storage]).

0.2::node(lifeSciences, 8, [video4], [access_logs, authentication, access_control, iot_data_encryption, firewall, host_IDS, pki, wireless_security, encrypted_storage]);
0.8::node(lifeSciences, 4, [video4], [access_logs, authentication, access_control, iot_data_encryption, firewall, host_IDS, pki, wireless_security, encrypted_storage]).

0.2::node(firePolice, 16, [video2, alarm1], [access_logs, access_control, authentication, backup, resource_monitoring, iot_data_encryption, firewall, host_IDS, pki, wireless_security, encrypted_storage]);
0.8::node(firePolice, 8, [video2, alarm1], [access_logs, access_control, authentication, backup,resource_monitoring, iot_data_encryption, firewall, host_IDS, pki, wireless_security, encrypted_storage]).

0.98::link(parkingServices, westEntry, 15, 70).
0.98::link(westEntry, parkingServices, 15, 70).
0.98::link(parkingServices, lifeSciences, 15, 70).
0.98::link(lifeSciences, parkingServices, 15, 70).
0.98::link(westEntry, firePolice, 15, 70).
0.98::link(firePolice, westEntry, 15, 70).
\end{lstlisting}

\noindent
\rev{where nodes feature different hardware capabilities\footnote{For the sake of the example, we limit the probabilistic variations to hardware capabilities. Of course, \prototypefn allows to associate to each probability a different node configuration also for what concerns IoT devices, that can temporarily fail or disconnect, and security countermeasures, that could be activated/deactivated depending on contextual factors, e.g. to save residual battery power. } according to a probability distribution and links are assumed to be wireless links with an associated reliability of $98\%$.}

\rev{Querying the predicate \texttt{placement(cctv_driver, P, R)} will output the following placements, the associated routing directives and a probability value ranking them as per how well they can satisfy all chain requirements: 
}

\vspace{1mm}
\begin{Verbatim}[fontfamily=courier, fontsize=\footnotesize, frame=single, framesep=1mm, framerule=0.1pt, rulecolor=\color{gray}]
placement(ucdavis_cctv,
 [on(cctv_driver,parkingServices), 
  on(feature_extr,firePolice), 
  on(lw_analytics,firePolice)],
  [(westEntry, firePolice, 15, [(cctv_driver, feature_extr)]), 
   (parkingServices, westEntry, 15, [(cctv_driver, feature_extr)])]): 0.9604

placement(ucdavis_cctv,
 [on(cctv_driver,parkingServices), 
  on(feature_extr,lifeSciences), 
  on(lw_analytics,lifeSciences)],
  [(parkingServices, lifeSciences, 15, [(cctv_driver, feature_extr)])]): 0.196
\end{Verbatim}

\noindent
\rev{where \texttt{cctv_driver} is placed on the \texttt{parkingServices} and \texttt{feature_extr} and \texttt{lw_analy-} \texttt{tics} can be placed both on the \texttt{firePolice} node or on the \texttt{lifeSciences} node, that now has a $20\%$ probability of featuring enough resources to host them. The best choice for the service chain deployer is still represented by the first output VNF chain placement and routing. It is worth noting that additional outputs with lower probability values could be kept as a backup deployments, when their associated probabilities exceed a desired threshold.}
\begin{flushright}
$\square$
\end{flushright}

\subsection{Complexity Analysis and Heuristics}
\label{sec:heuristics}

\rev{As per the considerations we have made in Sect.} \ref{sec:complexity_analysis}, \rev{the algorithmic time complexity of the approach described until now is exponential, i.e. $O(n^{s+{\tt Radius}})$. 
Besides, the probabilistic reasoning on disjoint clauses requires running} \prototype\ \rev{over an exponential number of possible worlds with a worst-case time complexity of $O(k^{n+m})$ where $k$ is the average number of disjoint clauses per each of the $n + m$ facts denoting the $n$ infrastructure nodes and the $m$ infrastructure links, respectively. For instance, in the case $k=2$, the overall complexity increases to $O(2^{n+m}\times n^{s+{\tt Radius}})$.}
Naturally, such time complexity becomes unbearable for very large infrastructures and for long service chains. Hence, we extended the prototype with a heuristic based on the probabilistic modelling we gave for the infrastructure \rev{capabilities}.

The heuristic version of \prototype allows users to specify two threshold values that are used to prune the search space whenever the probability of satisfying the chain hardware or bandwidth requirements, respectively, falls below them. Such pruning is implemented via the \problog\ subquerying system which is used to evaluate the probabilities of the {\tt servicePlacement} (i.e. \texttt{PHw}) and the {\tt path} (i.e. \texttt{PQoS}) goals during the search for eligible placements, and to check them against the user-specified thresholds (i.e. \texttt{THw} and \texttt{THw}). \rev{As soon as a candidate solution being built by the} \problog\ \rev{engine is associated to a probability lower than the user-set thresholds (and, thus, will not suitably satisfy the VNF chain requirements), the heuristic version of} \prototype\ \rev{stops searching along the corresponding path, hence reducing search times.} 

\rev{Particularly, the illustrated \textit{meta-reasoning} behaviour is implemented by simple extensions of the} \texttt{placement} \rev{and} \texttt{flowPlacement} \rev{predicates, as in } 
\begin{lstlisting}[]
placement(Chain, Placement, ServiceRoutes, THw, TQoS) :-
    chain(Chain, Services),
    subquery(servicePlacement(Services, Placement), PHw), 
    PHw >= THw, % meta-reasoning on servicePlacement/2
    flowPlacement(Placement, ServiceRoutes, TQoS).
    
flowPlacement(Placement, ServiceRoutes, TQoS) :-
    findall(flow(S1, S2, Br), flow(S1, S2, Br), ServiceFlows),
    flowPlacement(ServiceFlows, Placement, [], ServiceRoutes, [], S2S_latencies, TQoS),
    maxLatency(LChain, RequiredLatency),  
    latencyOK(LChain, RequiredLatency, S2S_latencies).

flowPlacement([],_,SRs,SRs,Lats,Lats,TQoS).
flowPlacement([flow(S1, S2, _)|SFs],P,SRs,NewSRs,Lats,NewLats,TQoS) :-
        subset([on(S1,N), on(S2,N)], P),
        flowPlacement(SFs,P,SRs,NewSRs,[(S1,S2,0)|Lats],NewLats,TQoS). 
flowPlacement([flow(S1, S2, Br)|SFs],P,SRs,NewSRs,Lats,NewLats,TQoS) :-
        subset([on(S1,N1), on(S2,N2)], P),
        N1 \== N2,
        subquery(path(N1, N2, 2, [], Path, 0, Lat), PQoS),
        PQoS >= TQoS, % meta-reasoning on path/7
        update(Path, Br, S1, S2, SRs, SR2s),
        flowPlacement(SFs,P,SR2s,NewSRs,[(S1,S2,Lat)|Lats],NewLats,TQoS). 
\end{lstlisting}

\rev{The new } \texttt{placement} \rev{predicate exploits} \problog\ \rev{built-in \texttt{subquery/2} to check whether historical variations in the behaviour of deployment nodes might lead to insufficient (hardware, IoT, security) resource availability for a certain placement of the services composing the VNF chain (lines 3--4). Similarly, the heuristic version of } \texttt{flowPlacement} \rev{ checks whether the historical variations in the behaviour of communication links might intolerably affect the routing of traffic flows along a certain path (lines 20--21).}

\medskip
\noindent\textbf{Example. } \rev{Running the heuristic prototype over the previous example and requiring that both deployment and communication requirements of output placement are met in $90\%$ of the cases, i.e. setting} ${\tt THw} = {\tt TQoS} = 0.9$, \rev{we only obtain as a result the first output placement which has an associated probability of $0.9604$.}
\begin{flushright}
$\square$
\end{flushright}

\section{EdgeUsher at Work}
\label{sec:experiments}

In this section, we illustrate a lifelike motivating example (Sect. \ref{sec:example}), and we discuss the \prototype prototype performance and the effectiveness of the proposed heuristics over such motivating example (Sect. \ref{sec:examplecontinued}).

\subsection{Motivating Example}
\label{sec:example}

\begin{figure*}[!h]
    \centering
    \includegraphics[width=.9\textwidth]{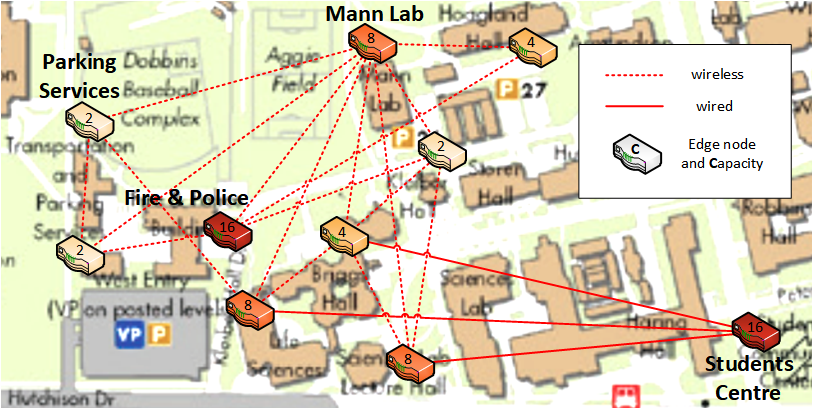}
    \caption{Example Edge infrastructure at UC Davis {\protect \cite{ning2019green}}.}
    \label{fig:ucdavis}
\end{figure*}

Hereinafter, we describe a lifelike example to better introduce the VNF placement problem and to highlight some of the related challenges. The example extends those that have been used throughout the paper to illustrate the proposed approach. 
We consider a portion of the topology of the Edge computing infrastructure deployed at UC Davis, inspired from \cite{ning2019green}, and sketched in Fig. \ref{fig:ucdavis}.
Such infrastructure is a Wireless-Optical Broadband Access Network (WOBAN) and consists of $10$ heterogeneously capable edge nodes.

\begin{figure}
\centering
{\footnotesize
\begin{tabular}{|c|c|c|}
\hline
\textbf{\thead{Node Type}} & {\thead{Hardware Profile}} & \textbf{\thead{Security Capabilities}} \\ \hline
    \makecell{\includegraphics[width=0.04\textwidth]{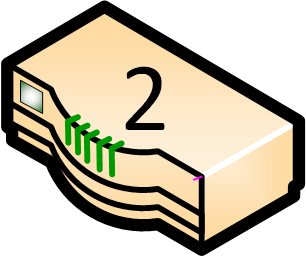}}      &  \makecell{20\% - 2; 80\% - 1}  &  \makecell{ authentication, anti-tampering,\\ wireless security, obfuscated storage  }    \\ \hline
    \makecell{\includegraphics[width=0.04\textwidth]{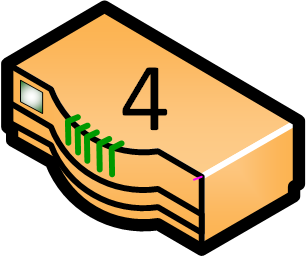}} &   \makecell{20\% - 4; 80\% - 2}                &     \makecell{ anti-tampering, authentication, IoT data encryption, firewall,\\ public key cryptography, wireless security, encrypted storage}                         \\ \hline
    \makecell{\includegraphics[width=0.04\textwidth]{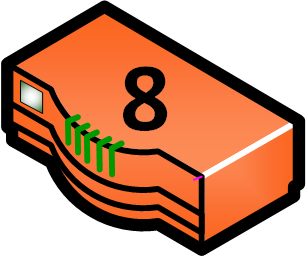}} &   \makecell{20\% - 8; 80\% - 4}               &    \makecell{ access control, access logs, authentication, IoT data encryption,\\ firewall, host IDS,public key cryptography,\\ wireless security, encrypted storage}                          \\ \hline
    \makecell{\includegraphics[width=0.04\textwidth]{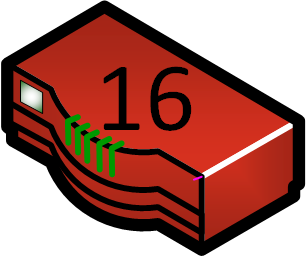}} &   \makecell{20\% - 16; 80\% - 8 }              &   \makecell{ access control, access logs, authentication, backup, \\resource monitoring,  IoT data encryption, firewall, host IDS,\\ public key cryptography, wireless security, encrypted storage}                           \\ \hline
    \makecell{ISP node} &   \makecell{20\% - 64; 80\% - 32 }            &    \makecell{ All}                          \\ \hline
    \makecell{Cloud} &   99.9\% - $\infty$               &   \makecell{ All}                           \\ 
    \hline
\end{tabular}
}
\vspace{2mm}
\caption{Example node types.}
\label{table:seccaps}
\end{figure}

We assume that available edge nodes feature either $2$, $4$, $8$ or $16$ hardware units\footnote{For the sake of readability, we only consider generic hardware units. Extensions to account for different resource types (e.g. RAM, CPU, storage) are straightforward.} and that they are subject to workload variations as per the distributions reported in Fig. \ref{table:seccaps}. For instance, nodes with 2 hardware units are totally free in $20\%$ of the cases, whilst they only have $1$ free hardware unit in the remaining $80\%$.
We also assume that different node types feature different security capabilities as reported in Fig. \ref{table:seccaps}, expressed in terms of a common vocabulary of edge computing security capabilities, as per the taxonomy of Fig. \ref{taxonomy}. 
Last, but not least, the nodes featuring $16$ GB of memory (viz., the \textit{Fire \& Police} and the \textit{Student Centre} devices) connect the edge network to a Cloud data centre through the same ISP Node (not shown in the figure).

Analogously, we assume that network links have the bandwidth and latency profiles listed in Fig. \ref{table:linkcaps}. For instance, on-campus wireless connections may be not available in $2\%$ of the cases, and feature 70 Mbps bandwidth and 15 ms latency in the remaining $98\%$.

\begin{figure*}[!ht]
\centering
{\footnotesize
\begin{tabular}{|c|c|}
\hline
\textbf{\thead{Link Type}} & {\thead{QoS Profile}}  \\ \hline
    \makecell{wireless (edge-edge)}    &       \makecell{98\% - 70 Mbps, 15 ms; 2\% - 0 Mbps, $\infty$ ms 
    }                                             \\ \hline
   \makecell{wired (edge-edge)}       &       \makecell{95\% - 250 Mbps, 5 ms; 5\% - 150 Mbps, 10 ms }                                            \\ \hline
   \makecell{wired (edge-ISP)}   &             \makecell{80\% - 1 Gbps, 10 ms; 20\% - 1 Gbps, 20 ms}                                        \\ \hline
    \makecell{Internet (ISP-cloud)}    &     \makecell{90\% - 10 Gbps, 50 ms; 10\% - 10 Gbps, 80 ms}                                               \\ \hline
\end{tabular}
}
\vspace{2mm}
\caption{Example link QoS profiles.}
\label{table:linkcaps}
\end{figure*}

We suppose that a new smart CCTV system has been installed at the \textit{Transportation \& Parking Services} building, and that it continuously captures video footage and streams it to a \textit{CCTV System Driver} deployed to the edge node which is in physical proximity.  

\begin{figure*}[!h]
    \centering
    \includegraphics[width=.9\textwidth]{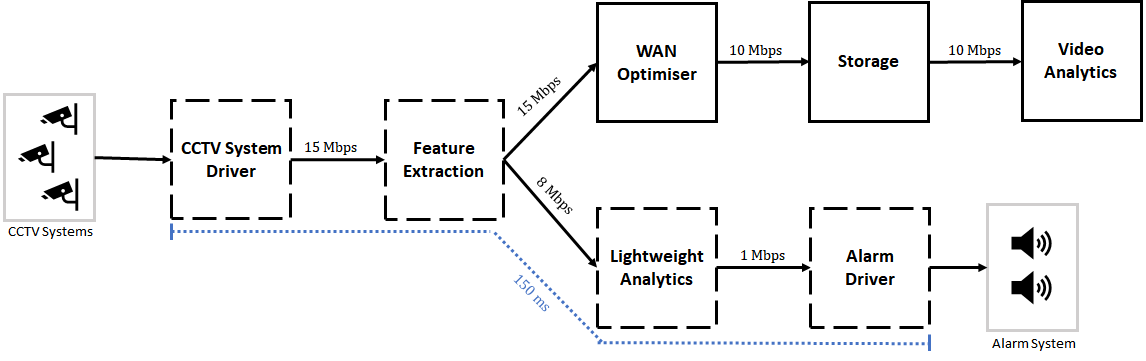}
    \caption{Example VNF chain.}
    \label{fig:request}
\end{figure*}

A VNF chain (Fig. \ref{fig:request}) must be deployed to support the video surveillance IoT system with a running application. The chain application, when suitably deployed, permits detecting events of interest (e.g., unauthorised access, fire, anomalous behaviour) by analysing video streams and by promptly notifying an alarm system installed at the \textit{Fire \& Police} station on campus. Such a VNF chain includes:
\vspace{-1.75mm}
\begin{itemize}
    \item a \textit{Feature Extraction} service function that applies image processing techniques to isolate potentially interesting video portions, and
    \item a \textit{Lightweight Analytics} service function that further processes such video portions by performing object recognition, by detecting anomalies or potentially dangerous situations, and by sending appropriate notifications to the \textit{Alarm Driver} deployed at the \textit{Police Station}.
\end{itemize}


\noindent To work as expected, the end-to-end latency from the {\it CCTV System Driver} to the {Alarm Driver} must not exceed $150$ ms latency, as shown in Fig. \ref{fig:request}. Additionally, for each link between two VNFs a minimum bandwidth requirement is specified, as shown in the figure. 
The traffic originated by the CCTV system is also collected  by a \textit{WAN Optimizer} service function that improves video data delivery efficiency (e.g., compression) and forwards video data to a {\it Storage} service. Complex video analytics are then performed with more relaxed latency constraints by a \textit{Video Analytics} service function which updates, when needed, the model used by the system to recognise potentially dangerous events.
Figure \ref{table:requirements} lists the requirements for the deployment of each VNF in terms of hardware units, connection to IoT devices (sensors or actuators) and security policies, along with the expected processing time of each chain function. As a further (soft) requirement, VNF chain deployers at UC Davis would prefer the \textit{Video Analytics} and \textit{Storage} service functions be placed on the same node (affinity) to reduce communication costs.

\begin{figure*}[!h]
\centering
{\footnotesize
\scalebox{0.95}{
\begin{tabular}{|c|c|c|c|}
\hline
\textbf{\thead{Service Function}} & {\thead{Hardware Requirements}} & \textbf{\thead{Security Requirements}} & \thead{Processing Time (ms)} \\ \hline
    \makecell{CCTV System Driver}      &  \makecell{1}  &  \makecell{anti-tampering $\vee$\\ access control}   &     \makecell{2}     \\ \hline
    \makecell{Feature Extraction} &   \makecell{3}                &     \makecell{access control $\wedge$ \\( obfuscated storage \\$\vee$ encrypted storage )}   &     \makecell{5}                         \\ \hline
    \makecell{Lightweight Analytics} &   \makecell{5}               &    \makecell{access control $\wedge$ host IDS $\wedge$ \\( obfuscated storage \\$\vee$ encrypted storage )}    &     \makecell{10}                         \\ \hline
    \makecell{Alarm Driver} &         \makecell{0.5}        &   \makecell{ access control $\wedge$ \\ host IDS }           &     \makecell{2}                   \\ \hline
    \makecell{Wan Optimiser} &    \makecell{10}           &    \makecell{ public key cryptography \\ $\wedge$ firewall $\wedge$ host IDS  }          &     \makecell{5}                   \\ \hline
    \makecell{Storage} &     \makecell{50}        &   \makecell{ backup $\wedge$ \\ public key cryptography }             &     \makecell{10}                 \\ \hline
    \makecell{Video Analytics} &      \makecell{16}           &   \makecell{resource monitoring $\wedge$ \\( obfuscated storage \\$\vee$ encrypted storage ) }            &     \makecell{ 40 }                  \\ \hline
\end{tabular}
}}
\vspace{2mm}
\caption{Example VNF requirements and processing times.}
\label{table:requirements}
\end{figure*}

\rev{Overall, both the available Cloud-Edge infrastructure and the VNF chain to be deployed on campus can be naturally declared in the input format exploited by} \prototype.
In fact, deploying the described chain to the infrastructure available at UC Davis implies solving the VNF placement problem, i.e. deciding how to map a VNF graph on top of an infrastructure substrate made of heterogeneous Edge and Cloud nodes and communication links, so that hardware, IoT and end-to-end network QoS requirements are all satisfied. 

Furthermore, the infrastructure is a dynamic environment and we assume it being subject to node workload variations and changing network conditions as per the probability distributions (possibly obtained from historical monitoring data \cite{forti2018mimicking}) we described in this section. 
Such changes can indeed affect deployment performance and turn momentarily optimal solutions into bad or unfeasible ones, potentially leading to unsatisfactory application QoS, and application downtime or unavailability.

As we will show in the next section, \prototype methodology permits determining VNF placement (i.e., function mappings and flow routes) and evaluating their performance against probabilistic infrastructure variations in this scenario.
In the next section, after discussing solutions to this first VNF placement, we will illustrate how the deployers at UC Davis can exploit our methodology to determine a further placement for the dashed part of the chain in Fig. \ref{fig:request}, handling a videostream for a second CCTV system deployed at the \textit{Mann Lab}, and joining the first chain at the {\it WAN Optimiser} service function.


\subsection{Motivating Example: Experiments}
\label{sec:examplecontinued}
\noindent We started by looking for eligible placements of the VNF chain supporting the CCTV system installed at the {\it Parking Services} building. 
For the purpose of the experiments\footnote{The experiments were run on a commodity laptop provided with an Intel Core i5-6200U CPU (2.30GHz) and 8GB of RAM, running Ubuntu 18.04.2 LTS,  \problog 2.1.0.36 and Python 3.6.}, we first run the non-heuristic \prototype over three different inputs (of the types (a)--(c) illustrated in Sect. \ref{sec:piedgeusher}) for the infrastructure description:

\vspace{-1.75mm}
\begin{itemize}
    \item [(a)] a \rev{\textit{non-probabilistic}} description that only considers the most probable values of each probability distribution \rev{without indicating the associated probabilities} (i.e., for both node hardware and link QoS profiles),
    \item[(b)] a \rev{\textit{single-probability}} description of the infrastructure that accounts only for the highest probability value of each distribution \rev{(i.e., one probability value per each node or link)}, and
    \item[(c)] a complete \rev{\textit{fully-probabilistic}} description of the infrastructure that includes all probability distributions available for nodes and links.
\end{itemize}

\noindent
\rev{Figure} \ref{fig:inputfiles} \rev{shows an example of the same link fact in the \textit{non-probabilistic}, \textit{single-probability} and \textit{probabilistic} infrastructure descriptions}\footnote{\rev{The three input files are available at:} \url{https://github.com/di-unipi-socc/EdgeUsher/tree/master/infra}}.

\begin{figure*}[!h]
\centering
\begin{lstlisting}[]
% non-probabilistic
link(isp, firePolice, 10, 1000).
% single-probability
0.8::link(isp, firePolice, 10, 1000).
% fully-probabilistic
0.8::link(isp, firePolice, 10, 1000);0.2::link(isp, firePolice, 20, 1000).
\end{lstlisting}
    \caption{\rev{Declaration of a \texttt{link} in the different input files.}}
    \label{fig:inputfiles}
\end{figure*}

\begin{figure}[!h]
\centering
\includegraphics[width=0.85\textwidth]{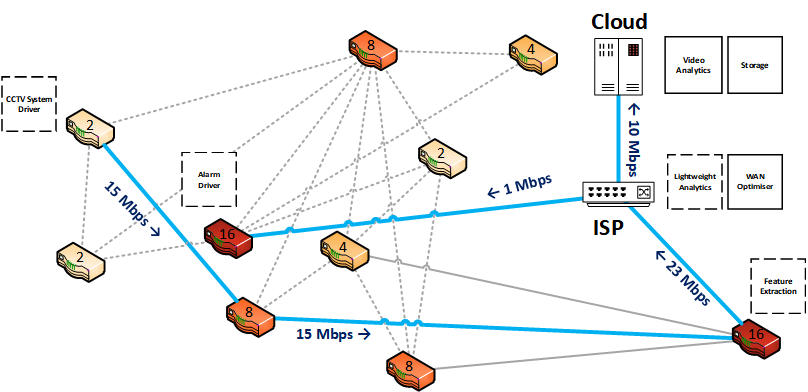}
    \caption{Best placement for the first chain \rev{(links are annotated with the bandwidth allocated to chain traffic flows)}.}
    \label{fig:deployment}
\end{figure}

\noindent Fig. \ref{table:resultsnonheuristic} shows the obtained results in terms of number of generated eligible placements and computation time\footnote{Timings obtained by averaging results over $15$ executions for each case.
} needed to obtain those. Fig. \ref{fig:deployment} shows one of the best placements obtained, which features $98\%$ probability of complying to all hardware, IoT, security, bandwidth and end-to-end latency requirements of the input chain.

\begin{figure*}[!h]
\centering
{\footnotesize
\begin{tabular}{c|c|c|}
\cline{2-3}
                                            & \textbf{\#Placements} & \textbf{Time} \\ \hline
\multicolumn{1}{|c|}{\textbf{Non-probabilistic}}     & 102                   & 8 s           \\ \hline
\multicolumn{1}{|c|}{\textbf{Single-probability}} & 102                   & 8 s           \\ \hline
\multicolumn{1}{|c|}{\textbf{Fully-probabilistic}}         & 4296                  & 2:48 h      \\ \hline
\end{tabular}
}
\vspace{1mm}
\caption{Results with \prototypefn without heuristics.}
\label{table:resultsnonheuristic}
\end{figure*}

It is worth noting that the prototype runs fairly fast on the non-probabilistic\footnote{The non-heuristic prototype can be run also in traditional Prolog environments to determine eligible placements in non-probabilistic infrastructure conditions. In this case, a first placement solution is returned istantaneously.} and on the single-probability infrastructure, which do not suffer from the additional combinatorial complexity that \problog incurs in, when evaluating (the probability distributions expressed as)  annotated disjunctions in the fully-probabilistic infrastructure description.
We will use the results obtained by the non-heuristic prototype over the fully-probabilistic description of the Edge infrastructure at UC Davis as a baseline to evaluate the performance of the heuristic prototype.

\begin{figure*}[!h]
\centering
{\footnotesize
\begin{tabular}{c|c|c|}
\cline{2-3}
                                            & \textbf{\#Placements} & \textbf{Time} \\ \hline
\multicolumn{1}{|c|}{\textbf{T=0.8}}         & 6                     & 47 s           \\ \hline
\multicolumn{1}{|c|}{\textbf{T=0.7}}         & 56                    & 56 s     \\ \hline
\multicolumn{1}{|c|}{\textbf{T=0.6}}         & 102                   & 1:07 min    \\ \hline
\multicolumn{1}{|c|}{\textbf{T=0.5}}         & 102                   & 1:07 min    \\ \hline
\multicolumn{1}{|c|}{\textbf{T=0.4}}         & 102                   & 1:13 min      \\ \hline
\multicolumn{1}{|c|}{\textbf{T=0.3}}         & 102                   & 1:22 min      \\ \hline
\multicolumn{1}{|c|}{\textbf{T=0.2}}         & 102                   & 1:37 min      \\ \hline
\multicolumn{1}{|c|}{\textbf{T=0.1}}         & 1056                  & 11:17 min      \\ \hline

\end{tabular}
}
\vspace{1mm}
\caption{Results with \prototypefn with heuristics.}
\label{table:resultsheuristic}
\end{figure*}

Thus, we run the heuristic \prototype over the fully-probabilistic description of the edge infrastructure at UC Davis. For the sake of simplicity, we set both thresholds (i.e., the one on node requirements {\tt THW} and the one on network QoS {\tt TQoS}) to a value $T$ that was varied during the experiments in the range $[0.1, 0.8]$ with a step of $0.1$. Figure \ref{table:resultsheuristic} shows the obtained results in terms of number of generated eligible placements and execution times$^{10}$ needed to obtain those.


The results show that the employed heuristics considerably reduces the search space and, thus, the execution time needed to determine eligible VNF placements and routing. Particularly, the fully probabilistic description of the UC Davis infrastructure can be handled in a time which shows a speed-up between $15$ and $214$ with respect to the exhaustive prototype, and still returns a subset of the optimal results.
Besides, with thresholds set to $0.8$, \prototype determines $6$ eligible placements for the VNF chain supporting the CCTV system. Four of such placement solutions have a probability of meeting all set requirements of $96\%$, whilst the remaining two of $98\%$. All output solutions fall within the best solutions generated also by the non-heuristic prototype, when it is run over the complete infrastructure description.

We then included an affinity constraint between the {\tt Storage} and the {\tt Video Analytics} service functions, and we run the prototype again \rev{over the fully-probabilistic input} with $T=0.8$. 
As a result, execution time halved with respect to the execution without such constraint, reaching around 24 seconds. The placement of Fig. \ref{fig:deployment} is still output as one of the best possible when forcing the {\it affinity} constraint. In this case, both the {\it Storage} and the {\it Video Analytics} are deployed to the available Cloud node. The highlighted links -- along with their labels -- show the routing path associated to the placement \rev{and the bandwidth to be allocated to traffic flows mapped on each infrastructure link}. Such piece of information could be actually used to instruct the network (e.g., via SDN controllers) so to allocate a suitable amount of bandwidth to each flow. 

\begin{figure*}[!h]
\subfloat[]{\includegraphics[width=0.85\textwidth]{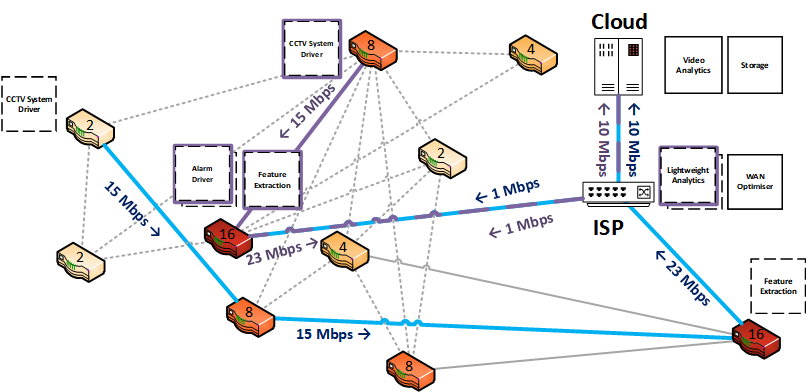}}\\
\subfloat[]{\includegraphics[width=0.85\textwidth]{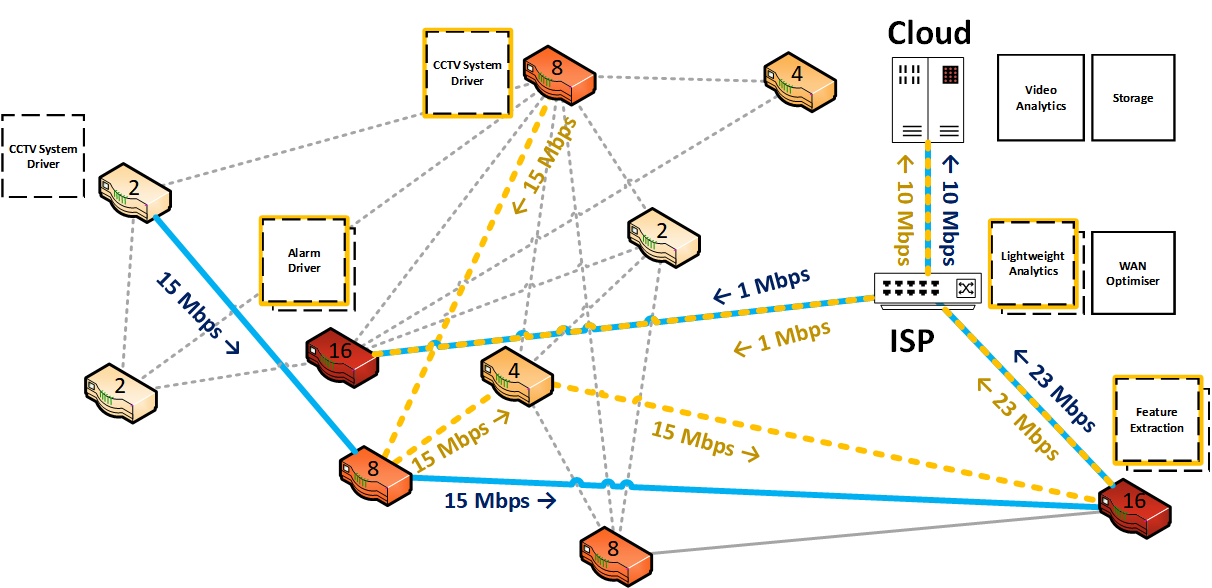}}
    
    \caption{Alternative placements for the second chain \rev{(links are annotated with the bandwidth allocated to chain traffic flows)}.}
    \label{fig:deployment2}
\end{figure*}

Afterwards, assuming the first chain was deployed as in Fig. \ref{fig:deployment}, \prototype was exploited to check whether it was possible to extend the deployment by placing anew the dashed part of the chain for a second CCTV system installed at the \textit{Mann Lab}. By querying again the heuristic prototype, $7$ new possible VNF placements were obtained in around $22$ s.
All output solutions featured a $96\%$ probability of meeting all chain constraints. Four of such solutions placed services as sketched in Fig. \ref{fig:deployment2} (a), the remaining three as in Fig. \ref{fig:deployment2} (b), other routings -- which are not shown -- were possible. It is worth noting that the deployers might consider using one of the output solutions and keep some of the others as possible backups to guarantee chain functioning in case of device failures or overloading, or in case of network congestion.

%


\section{Related Work}
\label{sec:related}


%


\rev{SDN and NFV technologies are gaining increasing interest for their potential benefits in hybrid Cloud-Edge environments} \cite{mouradian2018application} \rev{and in the IoT} \cite{Morabito:2017}. \rev{Indeed, the concept of Service Function Chaining (i.e., the ordered interconnection of service functions implemented as VNFs) is expected to enable the offer of \textit{added-value}~\textit{services}, like virtual reality or tactile Internet applications, over next-generation telecommunication networks} \cite{Cziva2018} \rev{and in Multi-access Edge Computing (MEC) scenarios} \cite{taleb2017multi,etsiMEC003}. 
%
Hereinafter, we discuss main related work in the area of SDN and NFV technologies applied in the IoT. 

An SDN and NFV architecture for IoT network and application management is proposed in \cite{ojo2016sdn}. 
\cite{Morabito:2017} propose an architecture and a prototype implementation of an NFV/SDN framework enabling automated and dynamic network service chaining across Edge (i.e., IoT gateway) and Cloud (i.e., central data center) domains. 
SDN and NFV are jointly used in \cite{fi9010008} to assure service continuity of a video monitoring application deployed over a flying ad-hoc network (FANET) built on a fleet of drones over rural areas. Drones are used as Point of Presence that can host Virtual Network or Application Functions.
 
The problem of placing VNFs on a physical substrate for realising service chains in a hybrid edge/cloud infrastructure to support IoT applications has only recently emerged.
Previous work has focused on network service placement in VNF infrastructure, considering intra and/or interDC networks \cite{pham2017traffic,LUIZELLI2017}. A survey on resource allocation strategies for the network services deployment in
VNF infrastructures has been provided by \cite{VNF_Herrera}.

Although converged approaches are emerging for managing NFV, edge and fog computing services \cite{commag_conv2017}, traditional VNF placement approaches do not tackle challenges brought by Fog and Edge computing for IoT applications. These challenges include heterogeneity of computing nodes, dynamic changes of network and node conditions that may turn optimal or quasi-optimal solutions into unfeasible ones, and security threats, just to mention the main ones. 
Recent work in the area of application placement in the Fog have partially begun to tackle these aspects, but open research problems still exist, such as placement approaches accounting for security aspects and dynamic infrastructure variations, as discussed in the review by \cite{DBLP:journals/corr/abs-1901-05717}.

Only few works have addressed the problem of placing VNFs in a hybrid environment made of edge and cloud computing nodes. 
\cite{Leivadeas2019} model the problem of SFC placement in hybrid MEC and cloud environment considering location requirements posed by VNFs and targeting minimisation of deployment costs and delays. They propose a Mixed Integer Programming (MIP) formulation of the problem and a sub-optimal approach based on the Tabu Search meta-heuristic.

SFC placement in IoT scenarios, which demand for low-latency response, high-throughput processing and cost effective resource usage, is tackled in \cite{Wang:IoT:2018}. The work proposes a linear programming model and an approximation optimization
algorithm to achieve deadline and packet rate guarantees while avoiding resource idleness. However, SFC orchestration is done within the cloud domain and the availability of computing resources at the edge is not considered.

\cite{Yala2018} propose a VNF placement algorithm that optimises access latency and service availability in a mixed edge and cloud environment for ultra-Reliable Low-Latency Communications (uRLLC) services. The multi-objective optimization problem is solved by exploiting a Genetic Algorithm metaheuristic, whose achieved performance is compared against an exact algorithm implemented in CPLEX. Although a network service is defined as a set of VNFs, chaining constraints are not considered.


\cite{mouradian2018application} tackle application component placement in NFV-based hybrid Cloud-Edge systems and propose an ILP formulation that represents applications as non-deterministic VNF Forwarding Graphs. Graphs can be built using sequence, parallel, selection and loop substructures and probabilities are used to model selection and loop iterations. 

Although all the above mentioned works \cite{Leivadeas2019,Wang:IoT:2018,Yala2018,mouradian2018application} consider latency requirements (either as minimisation objective or as constraint), none of them accounts for dynamic variations of network status, which instead can influence the extent to which QoS requirements are satisfied in the long run. Neither security aspects are taken into account. 

Typically, VNF placement approaches that consider dynamic network conditions either recompute placement \cite{Cziva2018}, enforce scaling and/or migration actions  \cite{eramo2017approach,jia2018online} or try to find a solution that is robust against network status variations \cite{Cheng_JSAC2018}.
\cite{Cziva2018} formulate the problem of edge VNF placement
as an ILP to derive latency-optimal deployments of VNFs. They also define a dynamic scheduler that recomputes placement to account for latency variations on links. This scheduling problem, which consists in  selecting the time for placement recalculation so that unnecessary VNF migrations are prevented and latency violations are bound, is solved using optimal stopping theory. \rev{While} \cite{Cziva2018} \rev{deals with placement of single VNFs and infrastructure dinamicity is modelled only in terms of network latency variation,} \prototype\ \rev{handles chains of VNFs and accounts for probabilistic distributions of latency and available bandwidth of links as well as of resource node capacity.}
In \cite{Cheng_JSAC2018} network dynamics are taken into account to find temporal robust placement solutions. 
The SFC placement is formulated as a Stochastic Resource Allocation problem that exploits both currently observed network information and future variation. However, the work does not tackle latency-aware placement and the network model does not represent variations of neither network latency nor node resources, as our work does.
\cite{EdgePlace2018} formulate a stochastic programming problem that minimizes the placement cost and aims at achieving high availability application deployments. The problem formulation thus accounts for probabilities of VM, host and link failures, but does not consider latency constraints.



 As analysed in \cite{Farris2019}, IoT environments introduce challenging security threats, ranging from attacks to IoT devices, attacks in IoT-oriented clouds and networks to threats in the application layer, such as vulnerabilities in software, data leakage and phishing. 
 Risks exist in 
 executing VNFs over third-party infrastructures and security and trust criteria have to inform placement decisions \cite{Farris2019}. Indeed, the need to consider security issues in virtual network embedding (VNE) and VNF placement problems is gaining increasing interest. A classification of security requirements into node, link and topological requirements to be considered in virtual network embedding problems is provided in \cite{Fischer:2017}.
In \cite{dwiardhika2019virtual} the problem of virtual network embedding is formulated so to account for standard protection provided by substrate nodes and links (quantitatively referred to as \textit{security level}). If the level of security is lower than the security demand, the VNE algorithm tries to place security VNFs (e.g.,  firewall, deep packet inspection, and intrusion detection) to improve the offered security level. Optimal placement of security SFCs is tackled in \cite{sendi2018}, where the placement problem is formulated including deployment constraints derived from network security patterns. 


Figure \ref{table:recap} provides a comparative overview of the discussed related work and highlights how, to the best of our knowledge, this is the first work aiming at addressing VNF chain placement in a hybrid edge/cloud network with latency constraints while accounting for network status variations and security requirements. In addition, while most related works rely on linear programming formulations, we adopt a probabilistic declarative approach. Indeed, declarative approaches have been successfully applied to modelling and reasoning on problems related to distributed systems other than VNF embedding -- as for instance in \cite{lopes2010applying} and \cite{ma2013declarative}.

\begin{figure*}[!h]
\centering
{\footnotesize
\scalebox{0.9}{
\begin{tabular}{|c|c|c|c|c|c|c|c|}
\hline
\thead{Article}   & \thead{\rot{Dynamicity}} &  \thead{\rot{Security}} & \thead{\rot{E2E Latency}} & \thead{\rot{Declarativity}} & \thead{\rot{Prototype}}  & \thead{\rot{Cloud-Edge}}  & \thead{\rot{IoT Devices}} \\ \hline

\cite{Leivadeas2019}  &  &   & \checkmark  &  &   & \checkmark & \checkmark\\ \hline

\cite{Wang:IoT:2018}  &  & & \checkmark     &  &   & & \checkmark\\ \hline

\cite{Yala2018}  &  &    & \checkmark    &  &   & \checkmark & \\ \hline

\cite{mouradian2018application}  &  &   & \checkmark    &  &   & \checkmark & \checkmark\\ \hline

\cite{Cziva2018}  & \checkmark &   & \checkmark    &  &   & \checkmark & \\ \hline

\cite{Cheng_JSAC2018}  & \checkmark &   &    &  &   &  & \\ \hline

 \cite{EdgePlace2018}  & \checkmark &    &    &  &   & \checkmark & \\ \hline

\cite{dwiardhika2019virtual}  &  & \checkmark  &    &  &   &  & \\ \hline

\cite{sendi2018} &  & \checkmark    &  &  &   &  & \\ \hline

{\prototypefn (this work)}  & \checkmark & \checkmark   & \checkmark  & \checkmark & \checkmark  & \checkmark & \checkmark\\ \hline

\end{tabular}
}
}
\vspace{2mm}
\caption{Related work overview.}
\label{table:recap}
\end{figure*}

Last but not least, our prototype is released as open source software and the experiment data is also made publicly available.

\section{Conclusions and Future Work}
\label{sec:conclusions}

In this article, we have proposed a logic programming approach for solving the problem of placing VNF chains onto Cloud-Edge infrastructures.
\ch{To achieve this objective, we followed three main steps:}
\begin{enumerate}
    \item \ch{we gave a concise logic programming formulation (hence, a declarative solution) of the considered VNF chain placement problem,}
    \item \ch{we extended it with a suitable probabilistic representation of variations in the available infrastructure capabilities to assess the quality of eligible solutions against those, and}
    \item \ch{we devised a heuristic to ensure scalability of our prototype and suitably high quality of output solutions, by immediately pruning out low quality solutions based on user-specified thresholds for hardware and QoS requirements.}
\end{enumerate}

\noindent
The obtained prototype, \prototype , returns the eligible deployments of VNF chains to a hybrid Cloud-Edge infrastructure that guarantee the fulfilment of a set of placement requirements, namely: hardware, IoT reachability, bandwidth, latency and security policies. Thanks to the declarative approach, additional constraints, such as affinity, anti-affinity or placement into a specific node can be easily expressed. Moreover, \prototype implementation has been fully provided  in this paper (around 70 source lines of code) together with some example of usage. \ch{Declarative programming also makes} \prototype\ \ch{ more flexible and extensible than procedural solutions, what makes it better suited to accommodate the ever-changing needs of Cloud-Edge scenarios.}

By leveraging \problog, \prototype 
permits specifying and solving the VNF chain placement problem 
while considering infrastructure variations, by assuming that probabilistic distributions describing the behaviour of computing nodes and infrastructure links are available. \ch{Since probabilistic logic programming is a natural extension of plain logic programming, it was straightforward to also consider dynamic infrastructure conditions by suitably extending (the input of) the Prolog version of} \prototype . \ch{Such an extension to account for dynamic settings would have required significantly more effort, if implemented by means of other paradigms.}
It is worth noting that \prototype could also be used to \rev{quickly} evaluate VNF placements computed by alternative placement algorithms with respect to the infrastructure variability, by calculating their probability of satisfying hardware and QoS requirements. 
In this work, we discussed and showcased the use of our prototype over a lifelike reference scenario, which we also used to assess and epitomise the performance of \prototype . 

\rev{Naturally, being the considered problem NP-hard, the worst-case time complexity of our approach is exponential in the size of the input infrastructure.}
For larger scenarios, we thus envision a hierarchical architecture of clusters of edge nodes (partitioned, for instance, according to administration, application or geographical criteria) where orchestration features are run by one head node, and connected to a few Cloud nodes. 
We intend to elaborate further on this vision by running \prototype over a domain made by a few clusters of edge nodes and their associated Cloud nodes.
\rev{However, the potential advantage of the probabilistic approach relies in the provisioning of solutions that are resilient to infrastructure variations over time. Our effort goes towards the direction of determining placements that are more likely to ensure high QoS-guarantees, security, and service reliability against dynamic infrastructure conditions, thus allowing amortising the cost of reasoning over an increased chain life time.} 

\rev{On direction for future work, we plan to comparatively evaluate our approach with state-of-the-art solutions that react to infrastructure changes by computing and executing costly VNF migrations or scaling actions (e.g.,} \cite{eramo2017approach,jia2018online}), \rev{using simulation as well as testbed environments}. The setting up of a small-scale testbed is undergoing in our campus network and it will be used to perform tests using probabilistic distributions derived from real monitoring data \cite{fogmon}. 

\prototype also allows specifying security requirements in terms of logical expressions over security properties. It is worth noting that some security properties can be provided exclusively as hardware capabilities (e.g., anti-tampering) while other ones could be implemented also as software and deployed as VNFs (e.g., firewall). This option opens up to the possibility of adaptively inserting required security VNFs when needed, which we also plan investigating in the near future.
\rev{Besides, we envision enhancing} \prototype\ \ch{with \textit{continuous} \textit{reasoning} capabilities}\cite{fogbrain} \ch{to further tame exp-time complexity at runtime}, automatic techniques to perform parameter tuning of the heuristic thresholds (e.g. via machine learning), and a user-friendly GUI to ease user interactions with the prototype.

\paragraph{Acknowledgements}
\noindent This work has been partly supported by the project
``{\it DECLWARE: Declarative methodologies of application design and deployment}"(PRA\_2018\_66), funded by University of Pisa, Italy, and by the project ``{\it GI\`O: a Fog computing testbed for research \& education}", funded by the Department of Computer Science of the University of Pisa, Italy.

\bibliographystyle{acm}
{\small
\bibliography{bibliography}}

\newpage
\appendix
\section{Code}

\subsection{{\textsf{EdgeUsher}} Prototype}

The complete code (72 sloc) of \prototype presented in Sect. \ref{sec:methodology} follows.

{\protect
\begin{lstlisting}
placement(Chain, Placement, ServiceRoutes) :-
    chain(Chain, Services),
    servicePlacement(Services, Placement, []),
    flowPlacement(Placement, ServiceRoutes).

servicePlacement([], [], _).
servicePlacement([S|Ss], [on(S,N)|P], AllocatedHW) :-
    service(S, _, HW_Reqs, Thing_Reqs, Sec_Reqs),
    node(N, HW_Caps, Thing_Caps, Sec_Caps),
    HW_Reqs =< HW_Caps,
    thingReqsOK(Thing_Reqs, Thing_Caps),
    secReqsOK(Sec_Reqs, Sec_Caps),
    hwReqsOK(HW_Reqs, HW_Caps, N, AllocatedHW, NewAllocatedHW),
    servicePlacement(Ss, P, NewAllocatedHW).

thingReqsOK(T_Reqs, T_Caps) :- subset(T_Reqs, T_Caps).

secReqsOK([],_).
secReqsOK([SR|SRs], Sec_Caps) :- subset([SR|SRs], Sec_Caps).
secReqsOK(and(P1,P2), Sec_Caps) :- secReqsOK(P1, Sec_Caps), secReqsOK(P2, Sec_Caps).
secReqsOK(or(P1,P2), Sec_Caps) :- secReqsOK(P1, Sec_Caps); secReqsOK(P2, Sec_Caps).
secReqsOK(P, Sec_Caps) :- atom(P), member(P, Sec_Caps).

hwReqsOK(HW_Reqs, _, N, [], [(N,HW_Reqs)]).
hwReqsOK(HW_Reqs, HW_Caps, N, [(N,A)|As], [(N,NewA)|As]) :-
    HW_Reqs + A =< HW_Caps, NewA is A + HW_Reqs.
hwReqsOK(HW_Reqs, HW_Caps, N, [(N1,A1)|As], [(N1,A1)|NewAs]) :-
    N \== N1, hwReqsOK(HW_Reqs, HW_Caps, N, As, NewAs).

flowPlacement(Placement, ServiceRoutes) :-
    findall(flow(S1, S2, Br), flow(S1, S2, Br), ServiceFlows),
    flowPlacement(ServiceFlows, Placement, [], ServiceRoutes, [], S2S_latencies),
    maxLatency(LChain, RequiredLatency),   %hp: only one maxLatency def
    latencyOK(LChain, RequiredLatency, S2S_latencies).

flowPlacement([], _, SRs, SRs, Lats, Lats).
flowPlacement([flow(S1, S2, _)|SFs], P, SRs, NewSRs, Lats, NewLats) :-
    subset([on(S1,N), on(S2,N)], P),
    flowPlacement(SFs, P, SRs, NewSRs, [(S1,S2,0)|Lats], NewLats). 
flowPlacement([flow(S1, S2, Br)|SFs], P, SRs, NewSRs, Lats, NewLats) :-
    subset([on(S1,N1), on(S2,N2)], P),
    N1 \== N2,
    path(N1, N2, 2, [], Path, 0, Lat),
    update(Path, Br, S1, S2, SRs, SR2s),
    flowPlacement(SFs, P, SR2s, NewSRs, [(S1,S2,Lat)|Lats], NewLats). 

path(N1, N2, Radius, Path, [(N1, N2, Bf)|Path], Lat, NewLat) :-
    Radius > 0,
    link(N1, N2, Lf, Bf),
    NewLat is Lat + Lf.

path(N1, N2, Radius, Path, NewPath, Lat, NewLat) :-
    Radius > 0,
    link(N1, N12, Lf, Bf), N12 \== N2, \+ member((N12,_,_,_), Path),
    NewRadius is Radius-1,
    Lat2 is Lat + Lf,
    path(N12, N2, NewRadius, [(N1, N12, Bf)|Path], NewPath, Lat2, NewLat).

update([],_,_,_,SRs,SRs).
update([(N1, N2, Bf)|Path], Br, S1, S2, SRs, NewSRs) :-
    updateOne((N1, N2, Bf), Br, S1, S2, SRs, SR2s),
    update(Path, Br, S1, S2, SR2s, NewSRs).

updateOne((N1, N2, Bf), Br, S1, S2, [], [(N1, N2, Br, [(S1,S2)])]) :-
    Br =< Bf.
updateOne((N1, N2, Bf), Br, S1, S2, [(N1, N2, Bass, S2Ss)|SR], [(N1, N2, NewBa, [(S1,S2)|S2Ss])|SR]) :- 
    Br =< Bf-Bass, NewBa is Br+Bass.
updateOne((N1, N2, Bf), Br, S1, S2, [(X, Y, Bass, S2Ss)|SR], [(X, Y, Bass, S2Ss)|NewSR]) :-
    (N1 \== X; N2 \== Y),
    updateOne((N1, N2, Bf), Br, S1, S2, SR, NewSR).

latencyOK(LChain, RequiredLatency, S2S_latencies) :-
    chainLatency(LChain, S2S_latencies, 0, ChainLatency),
    ChainLatency =< RequiredLatency.

chainLatency([S], _, Latency, NewLatency) :-
    service(S, S_Service_Time, _, _, _),
    NewLatency is Latency + S_Service_Time.
chainLatency([S1,S2|LChain], S2S_latencies, Latency, NewLatency) :-
    member((S1,S2,Lf), S2S_latencies),
    service(S1, S1_Service_Time, _, _, _),
    Latency2 is Latency+S1_Service_Time+Lf,
    chainLatency([S2|LChain], S2S_latencies, Latency2, NewLatency).
\end{lstlisting}
}

\subsection{Proof of Correctness and Termination of {\textsf{EdgeUsher}}}

We include here a sketch of the proofs of termination and correctness of \prototype.

\smallskip
\noindent
{\bf Proposition 1.} 
The query $placement(Chain,Placement,ServiceRoutes)$ always terminates.

\smallskip
\noindent
{\bf Proof.}
It is easy to prove that the query $placement(Chain,Placement,ServiceRoutes)$ always terminates since it:
\vspace{-0.2cm}
\begin{itemize}
\item
calls $chain$/2, which is matched against a set of facts and terminates immediately,
\item
calls $servicePlacement$/2 and $flowPlacement$/2, which both terminate.
\end{itemize}
\vspace{-0.2cm}
The call $servicePlacement(Services,Placement)$ terminates since:
\vspace{-0.2cm}
\begin{itemize}
\item
predicate $servicePlacement$/2 just calls $servicePlacement$/3,
\item
$servicePlacement$/3 performs tail-recursion by reducing the size of its first term (a list), so that if the size of the first term in the first call to $servicePlacement$/3 is $n$ then $servicePlacement$/3 performs $n$ tail-recursive calls and terminates,
\item
before tail-recurring, $servicePlacement$/3 
\vspace{-0.2cm}
\begin{itemize}
\item
calls $service$/5 and $node$/4, which are both matched against a set of facts and terminate immediately,
\item calls $thingReqsOK$/2, which scans $m$ times its second term (a list), where $m$ is the size of its first term (a list, too), and terminates,
\item
calls $secReqsOK$/2, which terminates 
\vspace{-0.2cm}
\begin{itemize}
\item 
either after scanning $m$ times its second term (a list), where $m$ is the size of its first term (if it is a list) 
\item 
or after recurring $m$ times by reducing the size of its first term (if it is an and-or term of depth $m$) and after scanning $m$ times its second term (a list),
\end{itemize}
\vspace{-0.2cm}
\item
calls $hwReqsOK$/5, which performs tail-recursion by reducing the size of its fourth term (a list), and terminates.
\end{itemize}
\end{itemize}

\vspace{-0.2cm}
The call $flowPlacement(Placement,ServiceRoutes)$ terminates since it:
\vspace{-0.2cm}
\begin{itemize}
\item
calls $findall$/3, whose inner goal is matched against a set of facts and terminates,
\item 
calls  $flowPlacement$/6, which performs tail-recursion by reducing the size of its first term (a list), and terminates;
before tail-recurring, $flowPlacement$/6 
\vspace{-0.2cm}
\begin{itemize}
\item 
calls $subset$/2, which scans twice its second term (a list),
\item 
calls $path$/7, which performs tail-recursion by reducing the size of its third term (a natural number), and terminates,
\item 
calls $update$/6, which performs tail-recursion by reducing the size of its first term (a list), and terminates,
\item 
before tail-recurring, $update$/6 calls $updateOne$/6, which performs tail-recursion by reducing the size of its fifth term (a list), and terminates
\end{itemize}
\vspace{-0.2cm}
\item calls $maxLatency$/2, which is matched against a set of facts and terminates immediately,
\item calls $latencyOK$/3, which just calls $chainLatency$/4
\begin{itemize}
\vspace{-0.2cm}
\item $chainLatency$/4 performs tail-recursion by reducing the size of its first term (a list), and terminates
\item before tail-recurring, $chainLatency$/4 calls $service$/5 (which is matched against a set of facts and terminates immediately) and scans once its second term (a list).
\hfill{$\diamond$}
\end{itemize}
\end{itemize}

\medskip
\noindent
{\bf Proposition 2.} 
If $servicePlacement([s_1,\ldots,s_k], P)$ is proved with computed answer substitution $P=[on(s_1,n_1),\ldots,on(s_k,n_h)]$, then the service placement defined by $P$ satisfies all the IoT, security and hardware requirements of $[s_1,\ldots,s_k]$.

\smallskip
\noindent
{\bf Proof.} 
We first prove ---by induction on the size of the first term of $servicePlacement([s_1,\ldots,s_k], P)$ --- that:
\vspace{-0.2cm}
\begin{tabbing}
(*) \= if 
  \= $servicePlacement([s_1,\dots,s_k],P)
  \rightarrow^*$
  \\
  \> \> $servicePlacement([],[on(s_1,n_1),\dots,on(s_k,n_h)],[(n_1,hw_1),\dots,(n_h,hw_h)])
  \rightarrow$
  \\
   \> \> $true$
 \\
 \> then $\forall j \in [1,h]: hw_j =  \sum_{on(s_i,n_j)} hw\_reqs(s_i)  \leq hw_\_caps(n_j)$

\\
{\em (Base case)}
Trivial since if 
$servicePlacement([s_1],Placement)
 \rightarrow^*$
 \\ 
 $servicePlacement([],[on(s_1,n_1)],[(n_1,hw_1)])
 \rightarrow true$
\\
then
$hw_1 =  hw\_reqs(s_1)  \leq hw\_caps(n_1)$, by lines 13 and 24 of the code in Appendix A.

\\
{\em (Inductive case)}
\\
If
  $servicePlacement([s_1,\dots,s_k,s_{k+1}],Placement)$\\
  $\rightarrow^*
  servicePlacement([s_{k+1}],[on(s_1,n_1),\dots,on(s_k,n_h)],[(n_1,hw_1),\dots,(n_h,hw_h)])$
  \\
  $\rightarrow^* servicePlacement([],[on(s_1,n_1),\dots,on(s_k,n_h),on(s_{k+1},n_{h+1})],[(n_1,hw_1),\dots,(n_h,hw_h),$\\
  $(n_{h+1},hw_{h+1})])$
  \\
  $\rightarrow true$
\\
where $n_{h+1} \not\in \{n_1,\dots,n_h\}$ 
then \\
\> $\forall j \in [1,h]:$ \= $hw_j =  \sum_{on(s_i,n_j)} hw\_reqs(s_i)  \leq hw_\_caps(n_j)$ by inductive hypothesis, 
\\
\> and $hw_{h+1} = hw\_reqs(s_{k+1}) \leq hw\_caps(n_{h+1)}$ by lines 13 and 24, 27, 28 of the code in Appendix A.
\\
If
  $servicePlacement([s_1,\dots,s_k,s_{k+1}],Placement)$\\
  $
  \rightarrow^* servicePlacement([s_{k+1}],[on(s_1,n_1),\dots,on(s_k,n_h)],[(n_1,hw_1),\dots,(n_h,hw_h)])$
\\
  $\rightarrow^* servicePlacement([],[on(s_1,n_1),\dots,on(s_k,n_h),on(s_{k+1},\overline{n})],[(n_1,hw'_1),\dots,(n_h,hw'_h)])
  \rightarrow true$
\\
where $\overline{n} \in \{n_1,\dots,n_h\}$ then \\

\>$\forall j \in [1,h]$:
$n_j \neq \overline{n} \ \Rightarrow  \ hw'_j = hw_j =  \sum_{on(s_i,n_j)} hw\_reqs(s_i)  \leq hw_\_caps(n_j)$ by inductive hypothesis,\\
and 
\\
\> $n_j = \overline{n} \ \Rightarrow \ hw'_j =  \sum_{on(s_i,n_j)} hw\_reqs(s_i)  \leq hw\_caps(n_j)$ 
\end{tabbing}

\noindent by inductive hypothesis and by lines 24---26 of the code in Appendix A.

\smallskip 
\noindent
We now prove that if 
  $servicePlacement([s_1,\dots,s_k],P)
  \rightarrow^* 
  true$ with computed answer substitution $P = [on(s_1,n_1),\dots,on(s_k,n_h)]$
then the service placement defined by $P$ satisfies all the IoT, security and hardware requirements of $[s_1,\ldots,s_k]$.

\noindent 
The proof is by induction on the size of the first term of $servicePlacement([s_1,\ldots,s_k], P)$.

\noindent
{\em (Base case)}
If 
  $servicePlacement([s_1],P)
  \rightarrow^* 
  true$ with computed answer substitution $P = [on(s_1,n_1)]$
 then $P$ satisfies the IoT, security and hardware requirements of $[s_1]$
 by lines 16 and 19---22 of the code in Appendix A, and by (*) respectively.

\noindent
{\em (Inductive case)}
If 
  $servicePlacement([s_1,\dots,s_k],P)
  \rightarrow^* 
  true$ with computed answer substitution $P = [on(s_1,n_1),\dots,on(s_k,n_h)]$
 then $P$ satisfies all the IoT, security and hardware requirements of $[s_1,\ldots,s_k]$
 by inductive hypothesis and by lines 16 and 19---22 of the code in Appendix A and by (*), respectively.
\hfill{$\diamond$}

\end{document}